 \documentclass{myaa}

 \usepackage{amsmath}
 \usepackage{graphics}

\long\def\jumpover#1{{}}

\def\approxgt{\,\raise2pt \hbox{$>$}\kern-8pt\lower2.pt\hbox{$\sim$}\,}
\def\approxlt{\,\raise2pt \hbox{$<$}\kern-8pt\lower2.pt\hbox{$\sim$}\,}

\def\grad{{\bf\nabla}}

\def\ah{{\scriptscriptstyle 1/2}}

\def\s3half{{\scriptscriptstyle {3\over 2}}}

\def\d{\partial}

\def \th{\thinspace}
\def \ngth{\negthinspace}

\def \Teff{{$T_{ef\!f} $}}

\def \Mo{{$M_\odot $}}
\def \Mom{{M_\odot }}
\def \Lo{{$L_\odot $}}

\def\Ra{{R{\scriptstyle a}}}
\def\Pe{{P\ngth{\scriptstyle e}}}
\def\Pr{{P\ngth{\scriptstyle r}}}
\def\Nus{{N\ngth{\scriptstyle u}}}

\begin{document}

 \myjournal{Submitted to Astronomy \& Astrophysics}
 \mydate{February 1998}

 \thesaurus{add }

 \title{Turbulent Convective Cepheid Models: Linear Properties}

 \author{P.A. Yecko \inst{1}, Z. Koll\'ath \inst{2} \& J. R. Buchler \inst{1}
 }

 \offprints{J. R. Buchler}

 \institute{$^1$Physics Department, University of Florida, Gainesville, FL
32611\\
 $^2$Konkoly Observatory, Budapest, HUNGARY
 }

 \date{}

 \maketitle

 \begin{abstract}

 A one-dimensional turbulent convection model in the form of a
time-dependent diffusion equation for the turbulent energy is
incorporated into our numerical pulsation code.
 The effect of turbulent convection on the structural rearrangement of
the static equilibrium star is taken into account, and the linear
eigenvalues (periods and growth-rates) for the {\sl complete turbulent
convective nonadiabatic} pulsations of the equilibrium models are
calculated.
 The linearized code allows us to perform efficiently a systematic
survey of the stability of Cepheid models for a wide range of
astrophysical parameters ($L$,$M$,\Teff) and the turbulent convective
parameters.

 % It is found that for low order modes, convective
 %timescales remain relatively fast and that the
 %contributions of the convective flux divergence, the eddy stresses and,
 %to a lesser extent, the turbulent diffusion all contribute to the
 %stabilization of pulsations.

 A sensitivity analysis of the properties of several sequences of
Galactic Cepheids is performed with the goal of using observational
constraints, such as the shape and location of the fundamental and first
overtone instability strips, to calibrate the turbulent convective input
parameters.  The locations of the red and blue edges of the fundamental
and first overtone instability strips and the overall strip widths are
largely determined by the parameters that determine the mixing length,
and strengths of the eddy viscosity and of the convective flux.  The
remaining parameters can be used for fine-tuning agreement with
observations.

 \end{abstract}

 \keywords{Stars: oscillations  \th Cepheids -- \th RR Lyrae,
 Turbulence,
 Convection,
 Hydrodynamics,
 Instabilities,\hfill\break
 Mixing~Length Theory,
 Methods: numerical}

\hyphenation{in-dep-end-ent}

% ----------------------------------
 %\jumpover{

 \section{Introduction}

  Cepheid studies have acknowledged, right from the beginning, that
convection should occur in Cepheid envelopes, and that they should alter
the properties of both the pulsations and the underlying equilibrium
models.  Since the Prandtl number characteristic of stellar material is
very small, $\Pr \sim O(10^{-9})$, and the Rayleigh numbers associated
with the partial ionization regions (PIR) are enormous, $\Ra$
$\approxgt$ $O(10^{12})$, stellar convection is far into the hard
turbulent regime (Krishnamurti \& Howard \cite{Ruby}).  The cooler
Cepheid stars are believed to have the more extensive convection zones,
leading to pulsational stability beyond the red edge of the standard
instability strip (IS).  Because of the estimated inefficiency of
convection and because of the difficulties of treating turbulent
convection, a lot of work has disregarded it altogether, the hope having
been that, except for providing a red edge, convection would play a
subdued role.  And purely radiative models have indeed provided a
reasonable overall agreement with observations.

  More recently it has become increasingly clear that an impasse has
been reached with radiative models, and that it is not possible to
improve the agreement with observations (e.g. Buchler \cite{Buchler97}).
In particular, the amplitudes of pulsation are systematically too large.
No consistent set of pseudo-viscous parameters can correctly limit the
amplitudes of {\sl both} fundamental and first overtone pulsations
without adversely affecting the stability of the corresponding limit
cycles, and the agreement with observation.  The new EROS and MACHO data
of the low metallicity Magellanic Cloud Cepheids (Buchler et
al. \cite{BKBG}) reveals further disagreement between stellar evolution
and radiative stellar pulsation models.  Clearly some additional
physical dissipation is missing in the radiative codes (see also
Kov\'acs \cite{Kovacs}), and turbulence and convection are the primary
suspects.

 Simulation of even the simplest turbulent convection (TC) problem is a
formidable numerical challenge; to adequately resolve the many time- and
length-scales of fully turbulent 3D convection at stellar Rayleigh and
Prandtl numbers is totally out of question, and astrophysicists have
been looking for 1D recipes to give an approximate, but acceptable
treatment of this phenomenon.  Thus the 1925 mixing length picture of
Prandtl was readily adapted to stellar envelopes by B\"ohm-Vitense, and
subsequently reformulated in many variations, all of which are
equivalent for time-independent problems (for an overview, see Baker
\cite{Baker}).  But time-independent MLT was never able to resolve the
red edge problem, 
mostly because it is time-dependent dissipation
introduced by eddy viscosity that provides a clear red edge, as we
will show in \S~\ref{limsection}.
Spiegel \cite{Ed63} first extended MLT to the
time-dependent and nonlocal cases.  The earliest attempts to include a
time-dependent mixing length model in pulsation (Gough \cite{Gough},
Unno \cite{Unno}) were not successful because they were too local in
space.  Subsequently, Gough's theory was extended to be less local, and
eventually was used in a full linearization of Solar modes (Balmforth
\cite{Neil}), analogous to what we have done here.  Unno's theory was
also developed (Castor \cite{Castor}) as a diffusion model, and
simplified to a single equation for the turbulent energy $e_t$ by
Stellingwerf (\cite{Swf82}).  More recent applications of this and
similar turbulent diffusion recipes have been made by Gonczi \& Osaki
\cite{Gonc-O}, Bono \& Stellingwerf \cite{BS}, Gehmeyr \& Winkler
\cite{Geh-W}, hereafter GW) and Kuhfuss (\cite{KuhfuB}).  Further
closely related models also appear in the literature, such as those of
Canuto (\cite{Canuto}) but have yet to be applied.

 All 1D model TC equations contain several dimensionless, order unity
parameters that are directly related to the physical quantities of the
model.  In principle, the validity of the 1D model equation can be
checked against experiments or against detailed 3D simulations of
turbulent convection, and values of the unknown parameters can be
extracted.  However, neither such experiments nor such simulations are
presently available.  We are obliged to calibrate these parameters
indirectly with the help of observational astronomical constraints, as
suggested more than a decade ago by Stellingwerf (\cite{Swf82}).  In the
best outcome, observational constraints could select one model in favor
of the others.  On the other hand we might discover that single equation
models for turbulent convection are insufficient, or that plumes
(e.g. Rieutord \& Zahn \cite{R-Zahn}) play an essential role in the
convective transport.

This programme has recently been initiated by incorporating these
recipes into hydrodynamic pulsation codes.  Fully nonlinear models have
been used to determine the stability properties of convective pulsators
(e.g. Bono \& Stellingwerf \cite{BS}, Bono \& Marconi \cite{BM97}), but
this approach is inefficient, and unsuitable when a large
number of models is required.  The approach we follow in this paper is
to compute the linear nonadiabatic properties directly by linearizing
the hydrodynamic equations about the equilibrium model and by then
computing the linear eigenvalues, a procedure which is relatively fast.
This allows us to perform an extensive survey of the sensitivity to the
TC parameters.

 \section{The Turbulent Convective (TC) Model Equations}

 We have adopted a TC model equation that consists of a single diffusion
equation for the turbulent energy and that includes the effects of the
turbulent stress, turbulent energy flux and turbulent energy production
on the mean momentum and energy; our model is essentially that used by
Gehmeyr (\cite{Geh92}).  In the following, we reproduce our PDE's for
the purpose of notation and for completeness.  In addition to $dr/dt =
u$ we have the momentum equation, modified to include the effects of
turbulent stress, that are cast here in the form of a turbulent pressure
and an eddy pressure
 \begin{equation}\label{mom}
 {du\over dt} = -4\pi r^2 {\partial\over\partial m}
 \left( p+p_t+p_\nu \right)  - {G M(r)\over r^2} \th .
 \end{equation}
 The pressure $p$ and specific energy $e$ represent the combined gas and
radiation quantities. 
 We use the specific internal energy equation in the form
 \begin{multline}\label{teq}
  c_v {d T\over dt} =
   -{\left(p+\Bigl({\partial e\over \partial v}\Bigr)_T\right)}
 {\partial (r^2 u)\over\partial m} \\
  - {\partial \over\partial m} \left[ r^2
 \left( F_c+ F_r\right)\right] + {\cal C} \, .
 \end{multline}
 The turbulent specific energy itself satisfies a diffusion equation of the
form:
 \begin{multline}\label{ete}
  {de_t\over dt} = - {\partial \over\partial m}\left[ r^2
 F_t\right]
   - (p_t+p_\nu)  {\partial (r^2 u)\over\partial m}
-{\cal C}\, ,
 \end{multline}\label{flux}
 where,
 \begin{equation}
   {\cal C}= \alpha_d\th {e_t^\ah \over\alpha_\Lambda H_p}
  \th\left(e_t - S_t- e_o\right) \th .
 \end{equation}
 Following GW we have included a negligible, nonzero background
turbulent energy, $e_o$, which is set equal to a small constant
($e_o=10^{4}$ erg cm$^{-3}$) value to avoid numerical problems.
 The turbulent and eddy pressures are given by:
 \begin{equation}\label{pt}
 p_t = \alpha_p\th\th \rho \th e_t \, ,
 \end{equation}
 \begin{equation}\label{pnu}
 p_\nu = - \alpha_\nu \th\th \alpha_\Lambda H_p\th\rho\th e_t^\ah {\d
u\over\d r} \, .
 \end{equation}
  The mixing length $\Lambda$ appears as the product
$\Lambda=\alpha_\Lambda H_p$ to emphasize that it is taken to be some
fraction $\alpha_\Lambda$ of the pressure scale height $H_p = {p r^2
/(\rho GM)}$.
   The convective and turbulent fluxes are:
 \begin{equation}\label{fc}
 F_c = \alpha_c\alpha_\Lambda\th \rho\th e_t^{\ah} \th  c_p T\th Y \, ,
 \end{equation}
 \begin{equation}\label{ft}
 F_t =-\alpha_t\th \alpha_\Lambda H_p \th e_t^{\ah}
 \th {\partial\th e_t \over\partial r} \, .
 \end{equation}
 The source of turbulence that appears in ${\cal C}$ is given by
 \begin{equation}\label{st}
  S_t = \alpha_s\th \alpha_\Lambda\th ( e_t p \beta T \th Y /
\rho)^{\ah} \, ,
 \end{equation}
 where the instability criterion has been embedded into the
convenient, dimensionless entropy gradient:
 \begin{equation}\label{Y}
 Y = \left[ -{H_p \over c_p} {\partial s \over \partial
r}\right]_{\scriptscriptstyle +} \, .
 \end{equation}
 Here $\beta$ is the thermal expansion coefficient, and all other
symbols have their usual meanings.

In the absence of turbulent diffusion, the formulations of Stellingwerf
(1982), of GW, and ours are equivalent; all three reduce to
time-dependent local MLT in this approximation.  With the diffusion term
included, however, the formulations are no longer identical.  As
Eq.~(\ref{fc}) shows we have taken the convective flux to be directly
proportional to the entropy gradient $ds/dr$, in the manner of Gehmeyr;
but in Eq.~(\ref{ete}) we have chosen to describe the turbulent energy
as in Stellingwerf.  While this would represent an inconsistency in
entirely local MLT, here it simply implies a different evaluation of the
correlation $\langle w'T'\rangle$ in the definition of the source term,
$S_t$.  In single equation models such as this, the absence of a dynamic
equation for $\langle w'T'\rangle$ necessitates such ad hoc evaluations.

 The TC recipe contains seven dimensionless parameters of order unity
denoted by $\alpha$'s.  It is important to test the sensitivity of the
results to variations of the $\alpha$'s in some reasonable range, say
$0.1\approxlt\alpha\approxlt10$.  However, these parameters also serve
as direct connections to the physical inputs of the model.  For example:
$\alpha_d$, in the absence of diffusion, allows us to adjust the
timescale of the turbulent energy growth and decay (GW);
$\alpha_\Lambda$ controls the mixing length as a function of the local
pressure scale height; $\alpha_\nu$ represents a constant eddy
viscosity; and $\alpha_p=(\gamma_t-1)$ is related to the adiabatic
compressibility index $\gamma_t$ of turbulent eddies.

 In our numerical study Eqs.~(\ref{mom})--(\ref{ete}) are differenced in
the spirit of the Fraley scheme (see Stellingwerf \cite{Swf74} and
Buchler, Koll\'ath \& Marom \cite{BKM}) using the Lagrangean mass as the
radial variable.

 % BEGIN FIGURE 1
 \begin{figure}
 \resizebox{8.7cm}{!}{\includegraphics{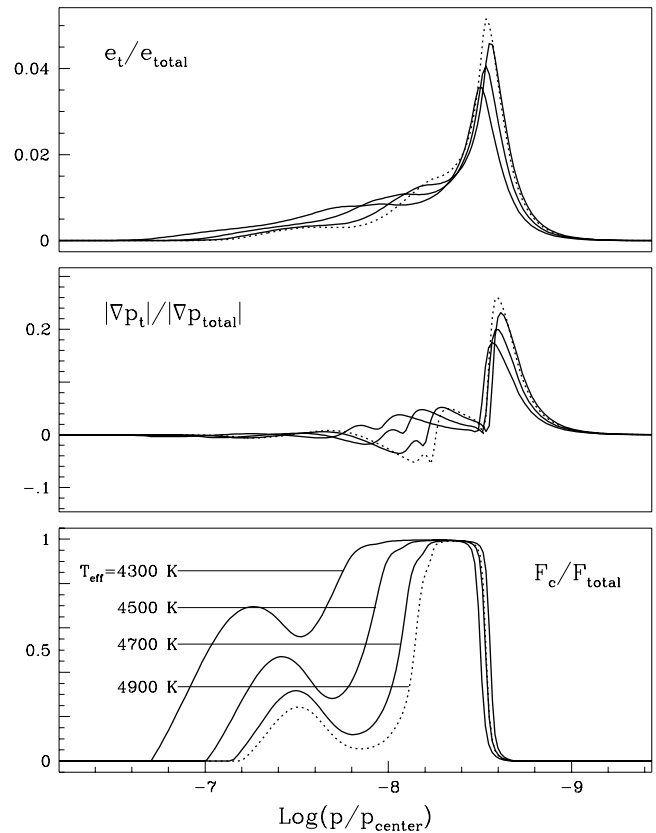}}
 \hfill
 \vskip 0pt
 \parbox[b]{8.7cm}{
   \caption{Standard Cepheid model sequence ($M$=5\Mo, $L$=2090\Lo,
   $\alpha_c$ = 3, $\alpha_\Lambda$ = 0.375)
   with \Teff (in K) = 4300, 4500, 4700, 4900;
   {\it top}: relative turbulent energy $e_t/e_{total}$;
   {\it middle}: relative turbulent pressure gradient $|\nabla p_t| /
|\nabla p_{total}|$,
 \th  ($p_{total}$=$p$+$p_t$); and
   {\it bottom}: relative convective flux $F_c/F_{total}$.
   }
 \label{teffig}
 }
 \end{figure}
 % END FIGURE 1

 % BEGIN FIGURE 2
 \begin{figure}
 \resizebox{8.7cm}{!}{\includegraphics{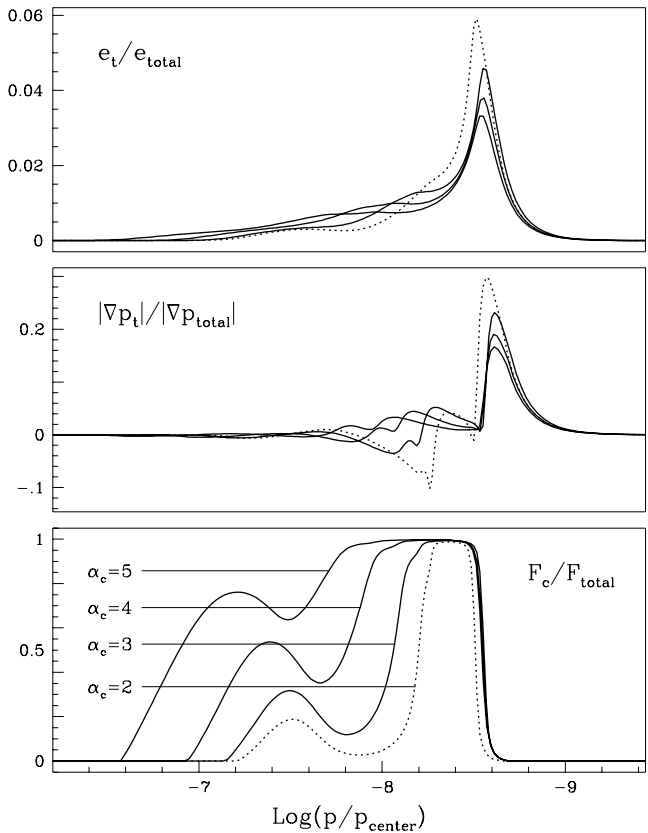}}
 \hfill
 \vskip 0pt
 \parbox[b]{8.7cm}{
   \caption{
   Effect of  $\alpha_c$ = 5,4,3,2;
 Cepheid models with $\alpha_\Lambda$ = 0.375, $M$=5\Mo, $L$ = 2090\Lo,
 and \Teff=4700\th K;
    {\it top}: $e_t/e_{total}$;
   {\it middle}: $|\nabla p_t| /| \nabla p_{total}|$; and
   {\it bottom}: $F_c/F_{total}$.
   }
 \label{acfig}
 }
 \end{figure}
 % END FIGURE 2

 \section{Static Models}

In Cepheids the PIRs are the source both of convection and of
pulsational driving.  Because convection causes a structural
rearrangement of the stellar model, convective stars can have different
properties from their radiative counterparts even if the direct
interplay between convection and pulsation is disregarded.  In our
linearized code it is possible to adjust parameters of either the
equilibrium model, the linearized model, or both; we can therefore show
the separate effects of the equilibrium structure and of the interaction
of convection and pulsation on the stability of the star.  We illustrate
this dichotomy in \S~\ref{limsection} by decoupling the turbulent and
hydrodynamic perturbations.  But first, we discuss the properties of the
static equilibrium models.

 % BEGIN FIGURE 3
 \begin{figure}
 \resizebox{8.7cm}{!}{\includegraphics{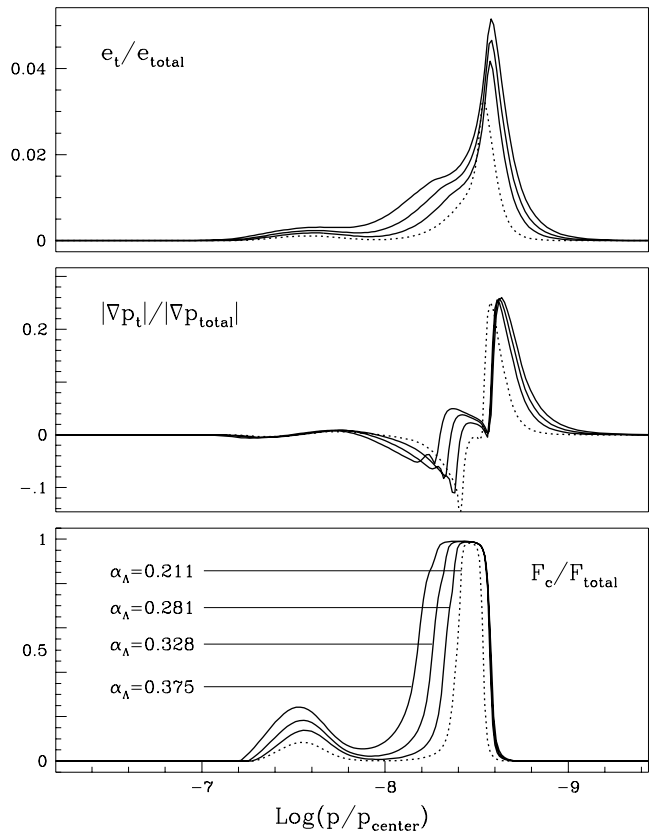}}
 \hfill
 \vskip 0pt
 \parbox[b]{8.7cm}{
   \caption{
   Effect of $\alpha_\Lambda$   = 0.375, 0.328, 0.281 and  0.211;
 Cepheid models with $\alpha_c$=3, $M$=5, $L$ = 2090, and \Teff=4700\th K;
   {\it top}: $e_t/e$;
   $|\nabla p_t| / |\nabla p_{total}|$; and
   {\it bottom:} $F_c/F_{total}$.
   }
 \label{amlt}
 }
 \end{figure}
 % END FIGURE 3

 The construction of the TC equilibrium models proceeds in two steps:
first, we integrate the time-independent form of
Eqs.~(\ref{mom}--\ref{ete}) from the surface inward, but in the
absence of turbulent diffusion ($\alpha_t$ =0).  This integration
becomes the initial guess from which we solve the complete
time-independent equation by Newton-Raphson iteration.  (As an
alternative we could iteratively solve Eqs.~(\ref{mom} -- \ref{teq})
followed by Eq.~\ref{ete}.)

In order to perform the extensive survey of this paper we had to use a
relatively coarse zoning of 180 mesh-points, viz. 40 uniform zones in
from the surface, then increasing geometrically up to an inner
temperature of 2.5\th million K.  Just as for the radiative models, it
is important to provide a grid point right in the center of the sharp
hydrogen PIR (unless a very fine mesh is used).  We accomplish this by
constraining the equilibrium model to have one anchor zone at
($T=11,000\th K$), holding this anchor fixed in the iteration to a full
TC model.  Failure to do so will result in erratic fluctuations of the
growth-rates along sequences of models.  We also find that for models
with strong convection which penetrates deeply into the stellar envelope
it becomes necessary to refine the zoning in the Fe~PIR, i.e. around
200,000\th K.

 We turn now to the properties of the equilibrium models.

\vskip -5pt

\subsection{The Convective Flux}

Along a Cepheid sequence (with fixed $M$, $L$ and composition) the size
of the negative entropy gradients increases when \Teff\ decreases,
causing convection to become more vigorous and to penetrate deeper into
the stellar envelope.  This is clearly illustrated in 
Fig.~\ref{teffig} for our standard sequence of models with decreasing
\Teff, for which we adopt the parameters $M$=5\Mo, $L$=2090\Lo, and with
\Teff=4300, 4500, 4700, 4900\th K, respectively.  Throughout this paper
we use the compositional parameters $X$=0.70, $Z$=0.02.  Similarly, for
the tests of this paper we adopt the {\sl standard} TC parameters,
viz. $\alpha_c$=5., $\alpha_t$=1., $\alpha_\nu$=3., $\alpha_\Lambda$ =0.375,
$\alpha_s$=0.75, $\alpha_d$=4., and $\alpha_p$=2/3.  These values are
just used for reference purposes and are not meant to be the values to
be adopted ultimately.  A detailed calibration of these parameters using
data from observations is in progress and will be addressed in a
subsequent publication.

In Fig.~\ref{teffig} (top) we display the fractional turbulent
energy $\th e_t/e_{total}\th $ for the model sequence, where
 $e_{total}=e+e_t$.
  The behavior of the corresponding turbulent pressure ratio $\th
p_t/p_{total}\th$ is similar, but slightly broadened because of the
$\rho$ dependence of $p_t$.  In the middle figure we show the relative
turbulent pressure gradient $|\grad p_t|/|\grad p_{total}|$.  Even
though the turbulent pressure generally does not exceed a few percent,
its gradient plays a more important role in the hydrostatic balance of
the convection zones.
 The bottom figure displays the fraction of total heat flux that is
carried by turbulent convection.  In the hydrogen PIR it is close to
100\%.

In Fig.~\ref{teffig} the temperature range 4300 -- 4900\th K corresponds
roughly to the width of the IS for these models; the coolest model, near
the red edge, is seen to have the most widespread turbulent pressure and
the deepest and most efficient domain of convective flux.  We point out
that it is the {\sl least} convective models that have the largest
turbulent pressure gradients.  This is an indirect result of those
models being less turbulent and consequently less diffusive.

The convection zones are of course associated with the PIRs.  In 
Fig.~\ref{teffig} the H, He$^+$, and He$^{\scriptscriptstyle ++}$
convection zones are all visible.  The maxima of the convective peaks
stay approximately at the same temperatures, but the structural changes
of the models with decreasing \Teff\ causes them to move to higher
pressures.  For the cooler model, the convection zone extends from the H
PIR all the way to the Fe peak PIR.

The convective flux $F_c$ \th (Eq.~\ref{fc}), depends on the parameter
$\alpha_c$ which we have held fixed in the sequence of
Fig.~\ref{teffig}.  We now consider the changes that occur when
$\alpha_c$ is increased for a model with fixed \Teff\ = 4700\th K and
with the same $M$ and $L$ as in Fig.~\ref{teffig}.  Interestingly,
Fig.~\ref{acfig} shows that a decrease in the strength of the convective
flux ($\alpha_c$) is remarkably similar to increasing the effective
temperature of the model.

\vskip -5pt

\subsection{Mixing Length}

The mixing length is the characteristic length-scale of the turbulent
eddies.  Therefore $\alpha_\Lambda$ appears in a number of places, and
in particular in the turbulent diffusivities, and  $\alpha_\Lambda$
represents a measure of the nonlocality of turbulence.  Its effect on
the spatial distribution of turbulent pressure and convective flux is
dramatic, and qualitatively different from that which results from
changes to $\alpha_c$ or \Teff.  In Fig.~\ref{amlt} we see that
the turbulent quantities energy (top), and pressure gradient (middle)
become very localized to the unstable regions even when $\alpha_\Lambda$
is made only slightly smaller than the standard value of
$\alpha_\Lambda$ = 0.375 that was adopted in
Figs.~\ref{teffig}-\ref{acfig}.  Concomitantly, the region of efficient
convective flux (Fig.~\ref{amlt}, bottom) separates into
increasingly narrow regions centered on the convectively unstable PIRs.

\vskip -5pt

\subsection{Properties of Convection}

\vskip -5pt

 \paragraph{Efficiency --}

\begin{figure}
 \resizebox{8.7cm}{!}{\includegraphics{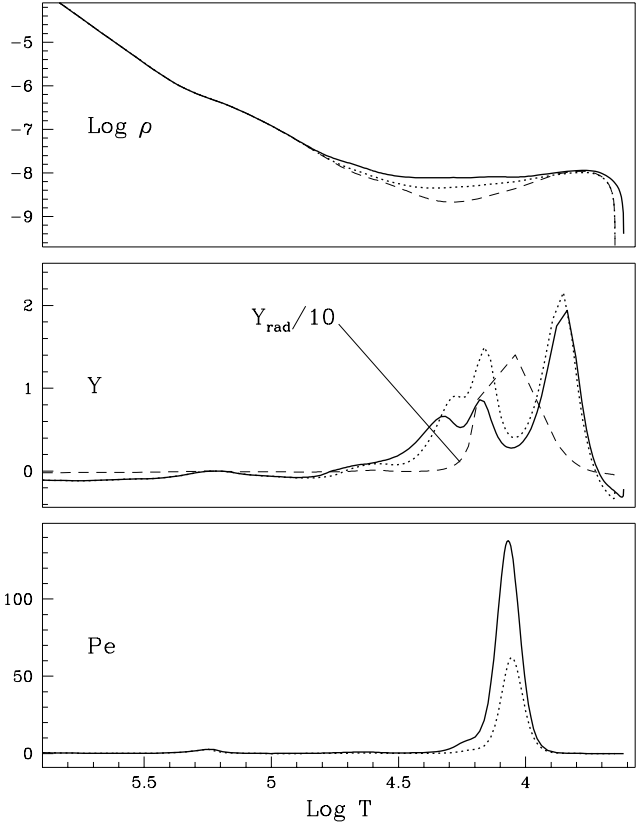}}
 \hfill
 \vskip 0pt
 \parbox[b]{8.7cm}{
   \caption{
 Structure of a typical Cepheid model ($M$=5\Mo, $L$=2090\Lo); 
  {\sl solid line:} \Teff=4900\th K; {\sl dotted line:} \Teff=5300\th K.
 The surface is on the right. All $\alpha$'s have standard values.
 {\sl Top:} density (Log) ;
 {\sl dashed line:} radiative model (\Teff= 5300\th K);
  {\sl middle:} dimensionless entropy gradient $Y$;
 {\sl bottom:} convective efficiency (P\'eclet number).
 }
 \label{structure}
 }
 \end{figure}

There are two conflicting definitions of convective efficiency.  The
P\'eclet number is one commonly used to measure the relative efficiency
of convective over conductive transport, and is given by $\Pe$ =
$t_d/t_c$, or the ratio of the thermal diffusion timescale to the
convective timescale.  Here, diffusion is radiative and its timescale is
given by $t_d \sim d^2 /\chi$, where $d=\alpha_\Lambda H_p$ is the
mixing length and $\chi = (16\sigma T^3)/( 3 c_p\rho^2\kappa)$ is the
usual radiative diffusivity.  The convective timescale is then given by
$t_c \sim d/e_t^{\scriptscriptstyle {1/2}}$.  The P\'eclet number for our
models is large in the convective region (bottom Fig.~\ref{structure}),
and convection is indeed efficient in the sense that it carries a large
fraction of the total heat flux in the PIRs
(Figs.~\ref{teffig} -- \ref{amlt}).

Truly efficient convection ought to remove the unstable gradient that
produces it.  In Fig.~\ref{structure}, top and middle, we show the
density and entropy profiles of two convective models.  We see that in
fact convection does not manage to remove the density inversion.  It
reduces the size of the negative entropy gradient by an order of
magnitude compared to the radiative model, but the remaining entropy
gradient is still strong, and it is shifted to the wings of the original
convection zone.

\vskip -5pt

 \paragraph{The Nusselt number $\Nus$ --}
 
This dimensionless number represents the relative importance of
convective and conductive fluxes, and is defined as $\Nus = F_c/F_{cond}$
for a system that is ordinarily conductive, but here it is radiative.
$\Nus$ is thus connected with the quantity plotted in Figs.~\ref{teffig}
-- \ref{amlt}.  

The Nusselt number must be related to the Rayleigh number, which is the
usual measure of the degree of convective instability, defined as $\Ra =
g \beta d^3 TY /( \nu \chi )$
 where $g = GM_r/r^2$ is the local gravity and $\nu$ is the kinematic
viscosity.  Thus $\Nus \sim \Ra^a$, and, under idealized conditions good
theoretical arguments have been given that, depending on the physical
assumptions (e.g. Spiegel \cite{SpiegelNu}), $a = 1/3$ or $a = 1/2$, but
for which no experimental support has yet been given.

\begin{figure}
 \resizebox{8.7cm}{!}{\includegraphics{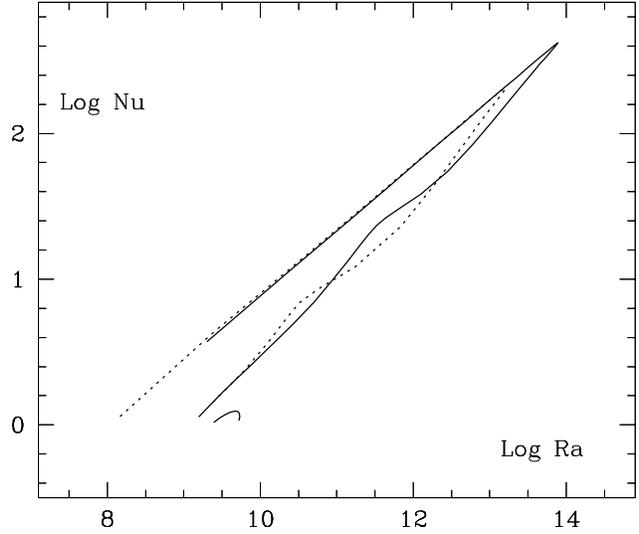}} \hfill
 \vskip 0pt
 \parbox[b]{8.7cm}{
 \caption{ Nusselt versus Rayleigh numbers throughout
 the convective regions (where $\Nus >$ 1) of two typical Cepheid envelopes
 with $M$=5\Mo, $L$=2090\Lo, \Teff=4900\th K ({\sl solid line}),
 \Teff=5300\th K ({\sl dotted line}).
 }
 \label{nusselt} }
 \end{figure}

 In Fig.~\ref{nusselt} we display the run of the local Nusselt number
versus the local Rayleigh number (assuming $\Pr = 10^{-9}$; this value
has essentially no effect on the slope) throughout two Cepheid
envelopes; the lines represent the same models that appear in
Fig.~\ref{structure}.  The slope lies between 0.45 and 0.53 throughout
the convective region, a value that agrees well with the theoretical
upper limit, considering the strong gradients and rapid variations of
the physical quantities (of $\chi$ among other things) in the convective
regions.

 % BEGIN FIGURE 4
 % END FIGURE 4

 % BEGIN FIGURE 5
 % END FIGURE 5

 \section{Linear Stability Analysis}

For the stability analysis we perform a linearization of all three
Eqs.~(\ref{mom}, \ref{teq}, \ref{ete}), with that of all their related
quantities Eqs.~(\ref{flux} -- \ref{Y}).
As pointed out earlier, the equilibrium model depends on the $\alpha$'s
(except $\alpha_\nu$), and consequently need to be reconstructed for any
change of $\alpha$ parameters before the linear problem is solved.

  Because of the turbulent energy equation (Eq.~\ref{ete}) there is an
additional spectral branch of eigenvalues compared to the vibrational
and thermal spectra of the nonadiabatic radiative models.  We {\sl
compute the complete spectrum of eigenvalues} by using canned eigenvalue
solvers (e.g. RG or EIGRF, as suggested by Glasner \& Buchler
\cite{Glasner}).  If only the lowest vibrational eigenvalues are desired
it is of course much faster to first compute the adiabatic eigenvalues
(symmetric band matrix) and insert these as initial guesses into a
Castor type iteration (Castor \cite{cas71}).

 % BEGIN FIGURE 6
 \begin{figure}
 \resizebox{8.7cm}{!}{\includegraphics{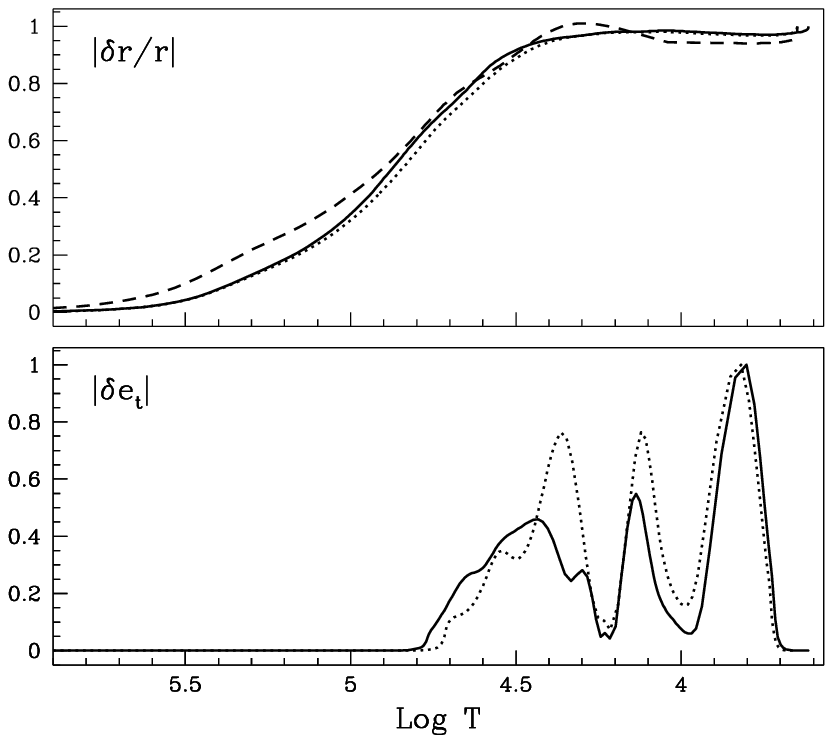}}
 \hfill
 \vskip 0pt
 \parbox[b]{8.7cm}{
   \caption{
 Fundamental eigenfunctions for the two Cepheid models with $M$=5, $L$=2090,
\Teff = 4900 (solid) and 5300\th K (dotted) as in Fig.~\ref{structure}.
 {\sl Top:} radial displacement vectors ($|\delta r|/r$);
 {\sl dashed line:} radiative model; 
 {\sl bottom:} turbulent energy eigenfunctions
 $|\delta e_t|$.
   }
 \label{eigvecs}
 }
 \end{figure}
 % END FIGURE 6

In this section we examine the role that the various ingredients of the
equations play in the stability of the low-lying vibrational modes of
Cepheid models.  The dominant effects are the nonlocality (mixing
length), and the strengths of the convective flux and of the eddy
viscosity, as we will show.

Cepheids are known to pulsate primarily in the fundamental or first
overtone modes, so we restrict our attention in the following analysis
to just these two modes.  The effects of turbulent convection on the
strange modes will be addressed in a companion paper.

\vskip -5pt

\subsection{Eigenfunctions}

Showing the effects of TC on all the lowest vibrational eigenvectors
would take us too far.  In Fig.~\ref{eigvecs} we limit ourselves to the
modulus of the (complex) radial displacement eigenvector $|\delta r/r|$
for two convective models (same as for Fig.~\ref{structure}).  The
purely radiative model (same $M$, $L$ and \Teff= 5300\th K) is shown as
a dashed line.  Convection removes some of the bump that occurs in the H
PIR because of the density inversion (cf. Fig.~\ref{structure}).

The bottom figure shows the Lagrangean perturbation of the specific
turbulent energy.  As expected $\delta e_t$ is localized to the convective regions.

 % BEGIN FIGURE 7
 \begin{figure}
 \resizebox{8.7cm}{!}{\includegraphics{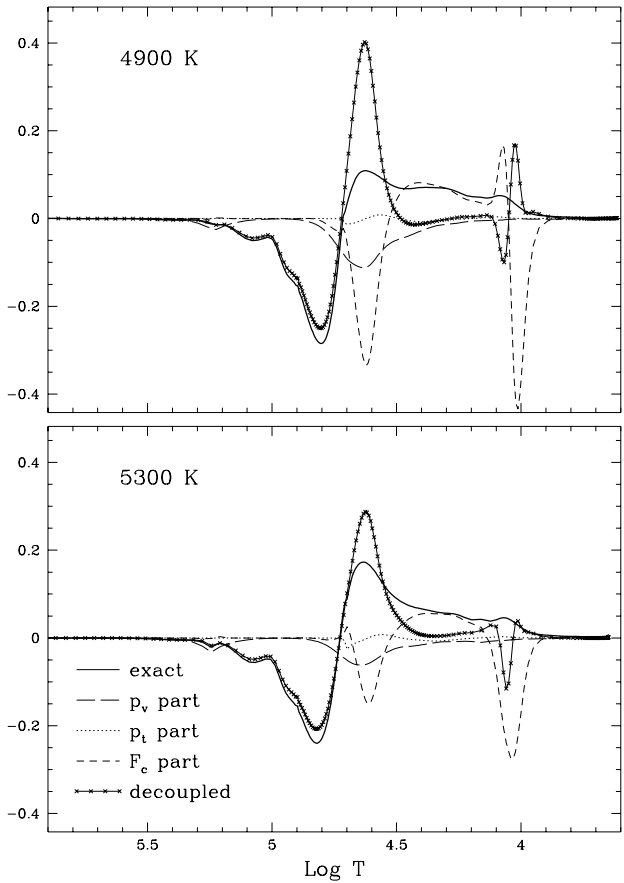}}
 \hfill
 \vskip 0pt
 \parbox[b]{8.7cm}{
   \caption{
    Work-integrand and its constituents for the models of Fig.~\ref{eigvecs}. 
 {\sl Top:} \Teff =4900\th K and {\sl bottom:} \Teff =5300\th K.
   }
 \label{works}
 }
 \end{figure}
 % END FIGURE 7

\subsection{Work-integrands}

In Fig.~\ref{works} we show the work-integrands\th ($\propto\delta
p_{total}\times\delta v^*$)\th for two models (top, \Teff =4900 and
bottom, \Teff =5300\th K) of Fig.~\ref{eigvecs}.  Shown are the exact
(or total) work-integrands (solid lines), together with the separate
contributions of the eddy pressure (long dashes) and the turbulent pressure
(dotted line).  The eddy pressure is always damping, as one can also
show theoretically.  The turbulent pressure can be both driving and
damping, but it always plays a subdued role in the overall growth-rate,
at least for the lowest frequency modes.

 The connected crosses show the decoupled case, i.e. the
work-integrands with the variations of all the TC quantities are
omitted.  Note that this is not equivalent to omitting both the $\delta
p_t$ and $\delta p_\nu$, because the $\delta p$ includes a contribution
from $\delta F_c$ and $\delta \cal{C}$; cf. Eq.~(\ref{teq}).

 The short-dashed line shows the work-integrand when the
variation $\delta F_c$ alone is omitted from  the linearization.

 % BEGIN FIGURE 8
 \begin{figure}
 \resizebox{8.7cm}{!}{\includegraphics{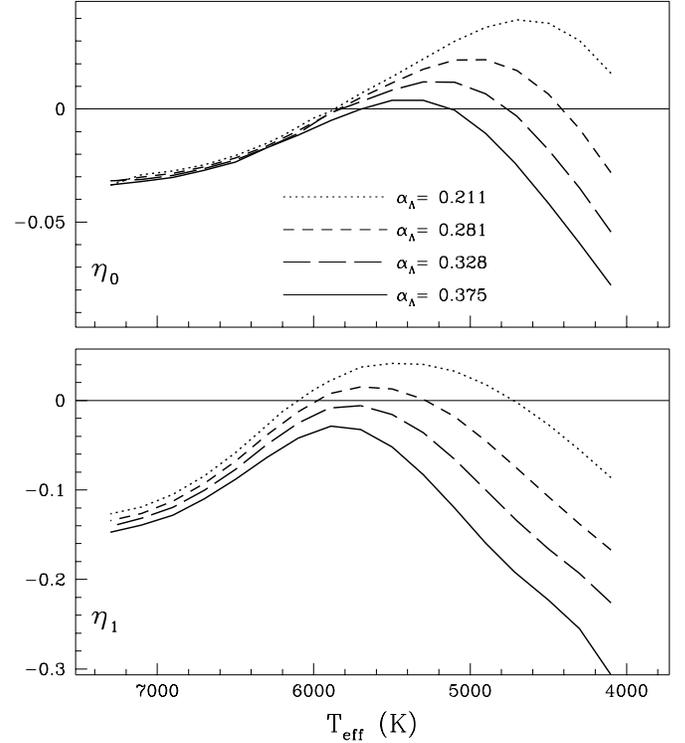}}
 \hfill
 \vskip 0pt
 \parbox[b]{8.7cm}{
   \caption{Relative growth-rates for a Cepheid with fixed $M$, $L$
  and \Teff\ in the range: 4100--7300\th K.
    {\it Top}: fundamental mode; {\it bottom}: first-overtone.
    Individual curves correspond
  to different values of $\alpha_\Lambda$ as indicated in the upper panel.
   }
 \label{kmlt}
 }
 \end{figure}
 % END FIGURE 8

 % BEGIN FIGURE 9
 \begin{figure}
 \resizebox{8.7cm}{!}{\includegraphics{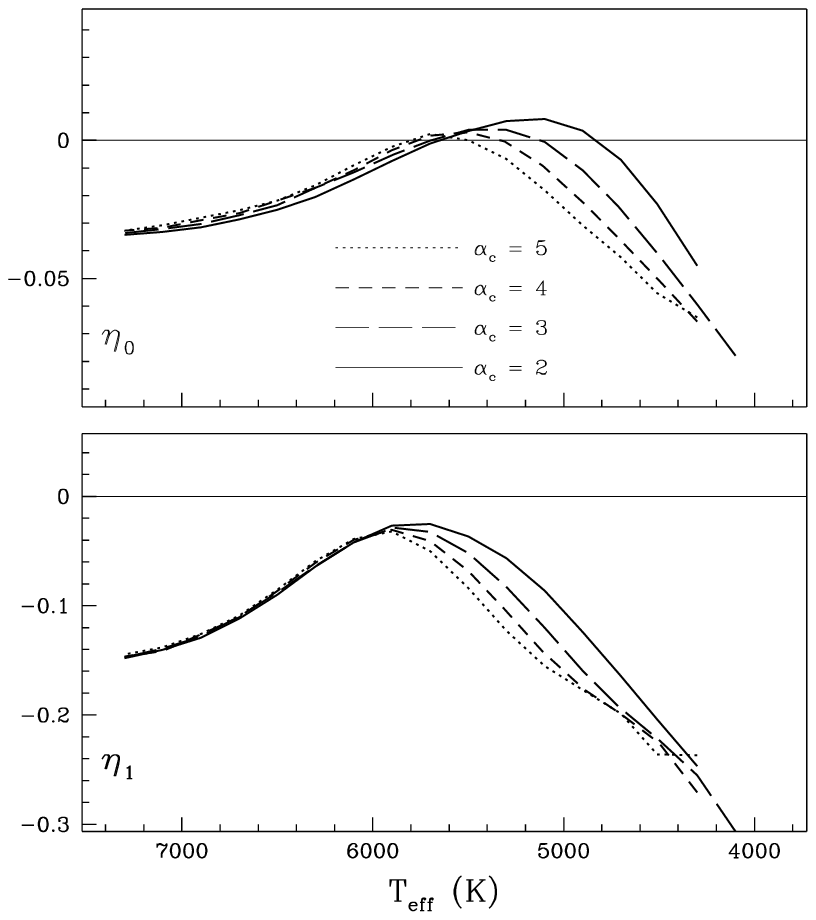}}
 \hfill
 \vskip 0pt
 \parbox[b]{8.7cm}{
   \caption{Relative growth-rates for a Cepheid with fixed $M$, $L$
  and \Teff\ in the range: 4100--7300\th K. 
   {\it Top}: fundamental mode, {\it bottom}: first-overtone.
   Individual curves correspond
  to different values of $\alpha_c$ as indicated in the upper panel.
   }
 \label{kc}
 }
 \end{figure}
 % END FIGURE 9

 % BEGIN FIGURE 10
 \begin{figure}
 \resizebox{8.7cm}{!}{\includegraphics{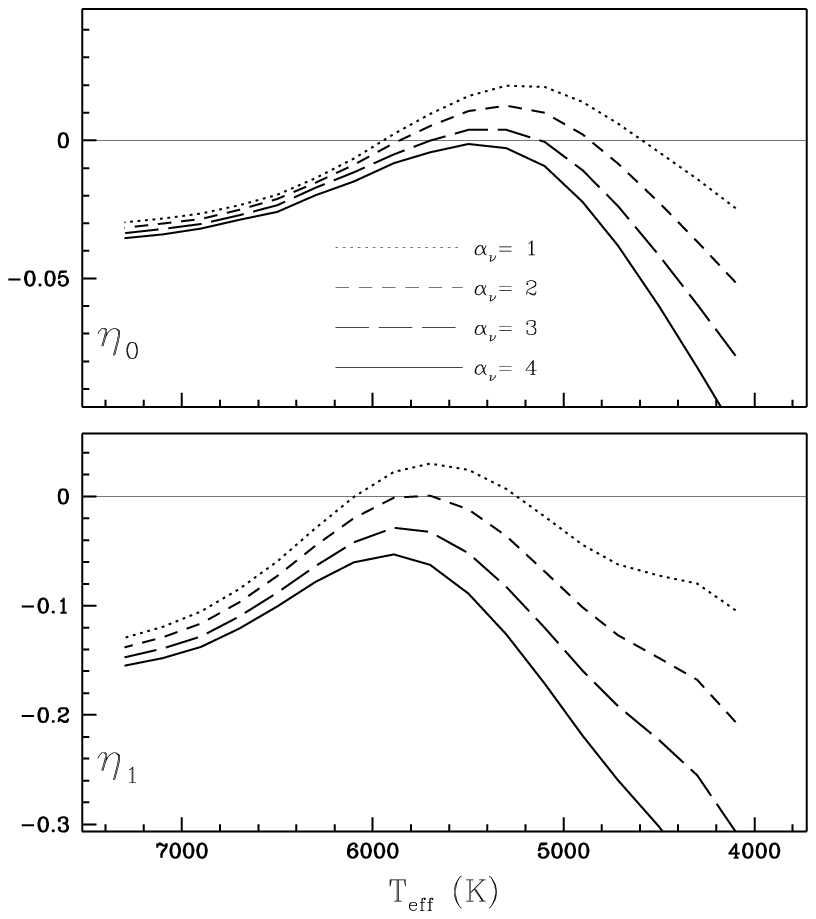}}
 \hfill
 \vskip 0pt
 \parbox[b]{8.7cm}{
   \caption{Relative growth-rates for a Cepheid with fixed $M$, $L$ and
\Teff\ in the range: 4100--7300\th K.
    {\it Top}: fundamental mode, {\it bottom}: first-overtone.
   Individual curves correspond
  to different values of $\alpha_\nu$ as indicated in the upper panel.
   }
 \label{kvt}
 }
 \end{figure}
 % END FIGURE 10

 \subsection{Mixing Length}

As we would expect from the behavior of the static models, the
properties of the pulsational modes are also found to be sensitive to
the value of the mixing length parameter $\alpha_\Lambda$.  In 
Fig.~\ref{kmlt} we show the growth-rates of the fundamental (top) and
first overtone (bottom) modes in four sequences each with temperature
ranging between $4000$ -- $7000$\th K, but having different
$\alpha_\Lambda$.  We define the relative growth-rates as those of the
model energies, i.e. $\eta_k=4\pi \kappa_k/\omega_k$, obtained from the
eigenvalues $\sigma_k = i\omega_k +\kappa_k$.

 % BEGIN FIGURE 11
 \begin{figure}
 \resizebox{8.7cm}{!}{\includegraphics{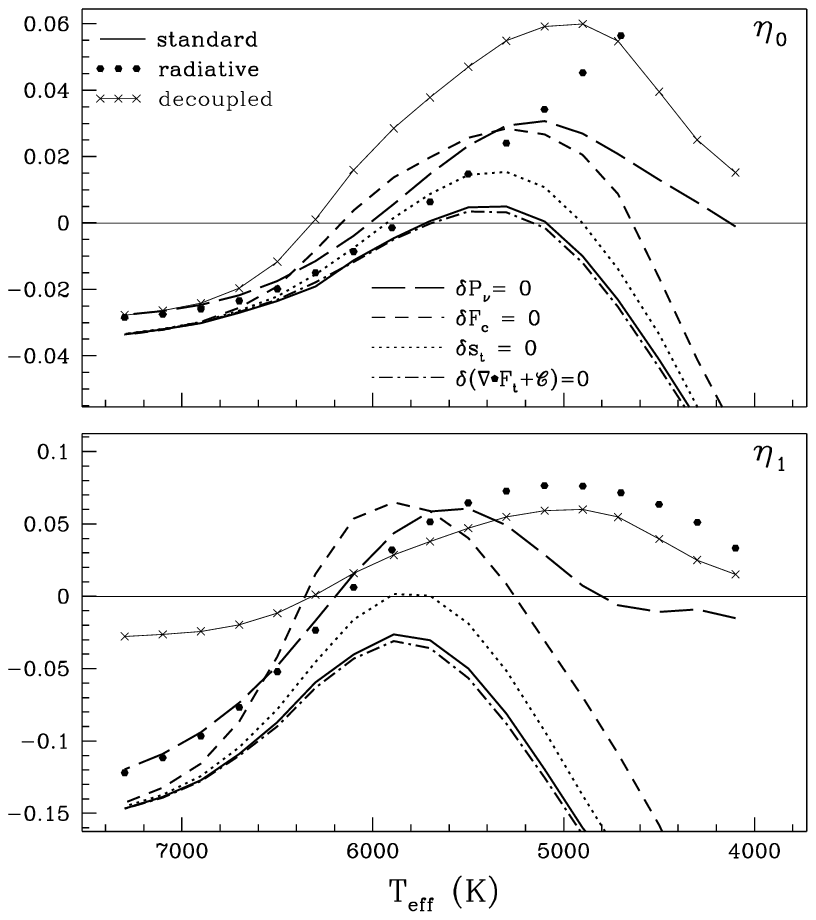}}
 \hfill
 \vskip 0pt
 \parbox[b]{8.7cm}{
   \caption{Fundamental and first overtone mode growth-rates for model
$M$=5\Mo, $L$=2060\Lo.
  {\sl Solid line:} exact values, other lines represent the various
  approximations indicated in the figure and described in the text;
  {\sl dots:} radiative model. 
 }
 \label{lims}
 }
 \end{figure}
 % END FIGURE 11

  The mixing length is the fundamental length-scale in TC models such as
this one.  The value of $\alpha_\Lambda$ therefore affects many of the
model quantities that have to be kept in mind when interpreting the
results of the variation of $\alpha_\Lambda$.  Here we perform our study
by varying {\sl only} $\alpha_\Lambda$ while holding {\sl all} other
$\alpha$'s fixed.

The most unstable sequence corresponds to the smallest value of the
mixing length (most local).

 \subsection{Convective flux}

While the behavior of the cooler models and of the higher frequency
modes is strongly tied to the mixing length the same is not true of the
convective flux parameter $\alpha_c$.  In Fig.~\ref{kc} we show
again the fundamental and first overtone growth-rates for the same star,
but now with $\alpha_\Lambda = 0.375$ and $\alpha_c = 2., 3., 4. ,5.$.

 We see here that while $\alpha_\Lambda$ has an enhanced effect on both the
cooler models and the overtone mode, $\alpha_c$ can affect the locations of
both the red edge and the blue edge {\sl and} the location of the maximum
growth rate.

 \subsection{Eddy Viscosity}\label{eddy-vis}

The eddy pressure represents the stress exerted on the fluid by the
turbulent eddies -- it can be characterized in terms of the eddy
viscosity parameter.  We can readily examine the effects of this term
on the damping of pulsational modes by varying the parameter
$\alpha_\nu$.  In Fig.~\ref{kvt} we show the fundamental and first
overtone growth-rates when $\alpha_\nu = 1.,2.,3.,4.$ ($\alpha_\nu
= 3.$ is what we take as the standard value).

 The effect on the overall shape of the IS is dramatic,
and for this reason, each of the parameter dependences above (that of
$\alpha_c$ and that of $\alpha_\Lambda$) is presented for a range of
eddy viscosities $\alpha_\nu$ = 1., 2., 3., 4.\th .

 % }
% ----------------------------------

\subsection{Limits: Adiabatic Turbulence and  Instantaneous
Turbulence}\label{limsection}

Numerically we also have the ability to disallow perturbations to
individual turbulent quantities; this is useful for determining which TC
components, when perturbed, play the largest role in the growth-rates,
for example.  We now use this tool to examine some limits for
the coupled convection and pulsation problem for our model with $M$=5,
$L$=2060.  Of particular interest are the adiabatic turbulent and the
instantaneous turbulent limits.

We can constrain the {\sl turbulent perturbations to be adiabatic} in
that the {\sl specific} entropy of the turbulent eddies remains
unchanged, i.e. $\delta s_t = \delta e_t -p_t\delta v =0$.  Note that
this is not equivalent to ignoring the perturbation of $\delta e_t$, nor
is it equivalent to assuming a very long TC timescale.  With the help of
Eq.~\ref{pt} one sees that this limit allows for the finite
compressibility of the turbulent energy, $e_t \sim \rho^{\alpha_p}$, or
$\gamma_t=1+\alpha_p$).  (We stress that our eigenvalues are fully
nonadiabatic though, as far as the usual gas and radiation entropies are
concerned).  These {\sl adiabatic turbulent} growth-rates can be seen in
the dotted curve of Fig.~\ref{lims} compared to the full nonadiabatic
turbulent growth-rates (solid line); the effect on these lowest modes is
not negligible showing 'adiabatic turbulence' to be an unacceptable
approximation.

At the other extreme, we can allow the {\sl turbulent energy
perturbations to respond instantaneously}, i.e. the turbulent timescale
is taken to be much shorter than the pulsation time.  In Eq.~\ref{ete}
this corresponds to setting the Lagrangean variation of the flux and
source terms equal to zero (note that $p_t$ is part of the turbulent
entropy, and $p_\nu$ disappears in the linearization).
 These {\sl instantaneous turbulent} growth-rates, as seen in the
bottommost curves of Fig.~\ref{lims}, lie very close to the exact
ones.  This indicates that overall, and at least for the fundamental
mode and first overtone, the turbulent convection response-times remain
much shorter than the period of the pulsation.  (If one wanted to
simplify the linear stability analysis for the lowest modes in Cepheid
models, one could thus drop the time-dependence in Eq.~\ref{ete} and
express $\delta e_t$ in terms of the variations of the other
quantities).

It is also interesting to ask which of the perturbed TC quantities
contribute most significantly to the overall driving and damping of the
pulsational modes.  We have therefore added four more curves in 
Fig.~\ref{lims}.\hfill\break
 1. The long-dashed curves show the growth-rates with the neglect of the
eddy viscosity perturbation ($\delta p_\nu $=0).  Without this critical
stabilizing contribution the width of the IS would be excessively large.
Comparison of this curve with those in Fig.~\ref{kvt} clearly shows that
as the eddy viscosity is reduced, the red edge in particular is rapidly
shifted to unrealistically low temperatures.
 \hfill\break
 2. The short-dashed curves represent the growth-rates with the neglect
of the Lagrangean convective flux perturbation ($\delta F_c $=0).  This
term is seen to play an equally important stabilizing role.
 \hfill\break
 3. The connected crosses show the growth-rates corresponding to the
neglect of the perturbations of {\sl all} TC quantities.  The effect of
turbulent convection appears only in the structure of the static model.
Not only are modes much less stable overall, in addition, the sequence
does not have a red edge within a reasonable \Teff.
 \hfill\break
 4. The large heavy dots represent a purely radiative sequence,
displaying a continuing instability for the coolest models (no red
edge with a reasonable \Teff).

 % BEGIN FIGURE 12
 \begin{figure}
 \resizebox{8.7cm}{!}{\includegraphics{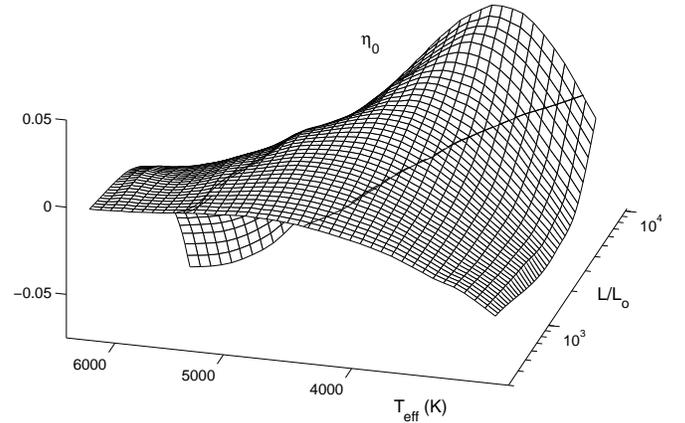}}
 \hfill
 \vskip 0pt
 \parbox[b]{8.7cm}{
   \caption{The interpolated data surface, here for the fundamental
  mode for $M/\Mom$=3, 4, 5, 6, 6.5, 7 (for $M$--$L$ relation, cf text)
 from which each IS is determined.  The $\eta_0$=0 contour superposed
 on the surface corresponds to the red (right) and blue (left) edges.
 }
 \label{etasurf}
 }
 \end{figure}
 % END FIGURE 12

 % BEGIN FIGURE 13
 \begin{figure*}
 \resizebox{\hsize}{!}{\includegraphics{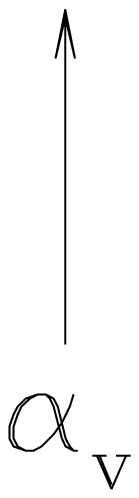}
                       \includegraphics{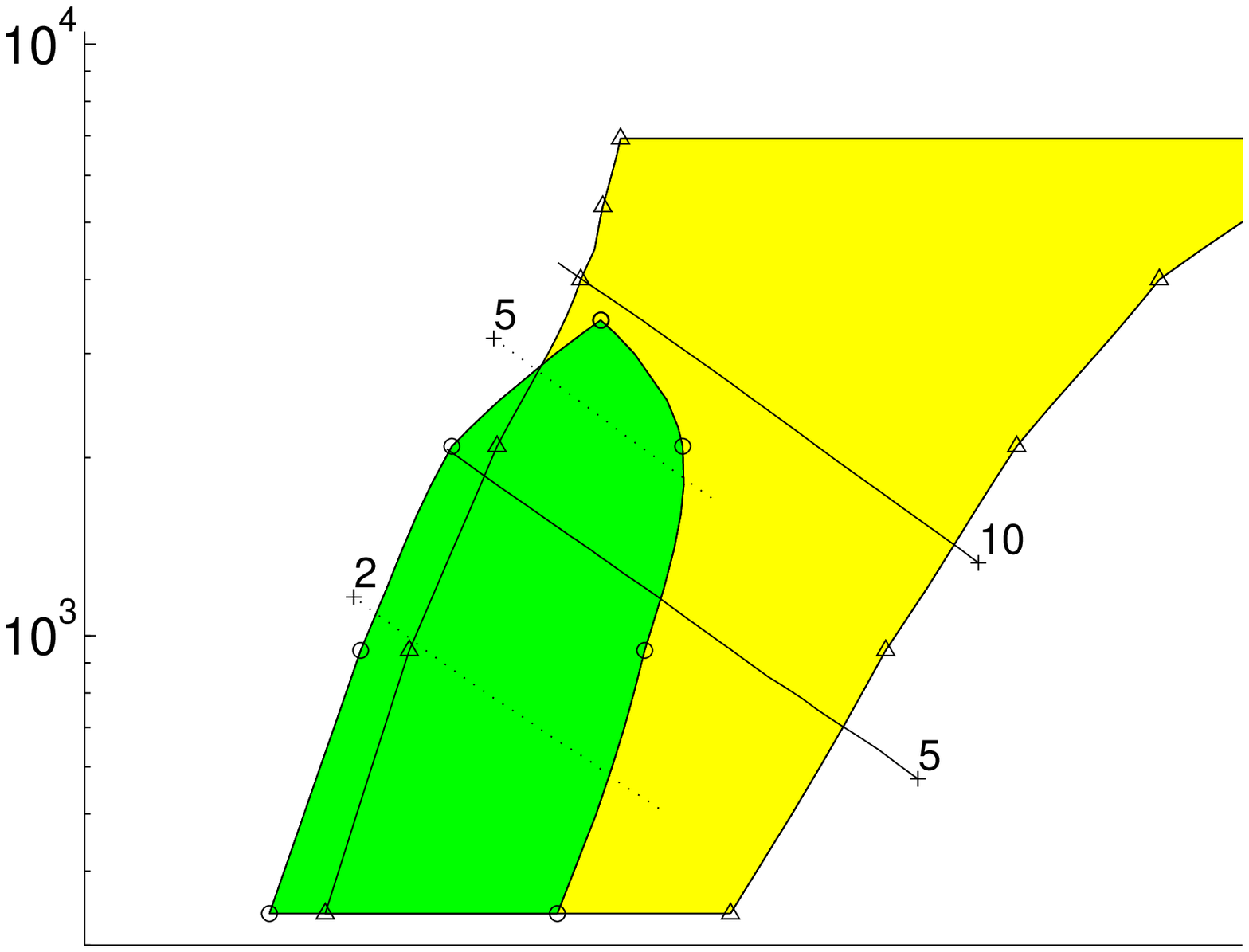}
                       \includegraphics{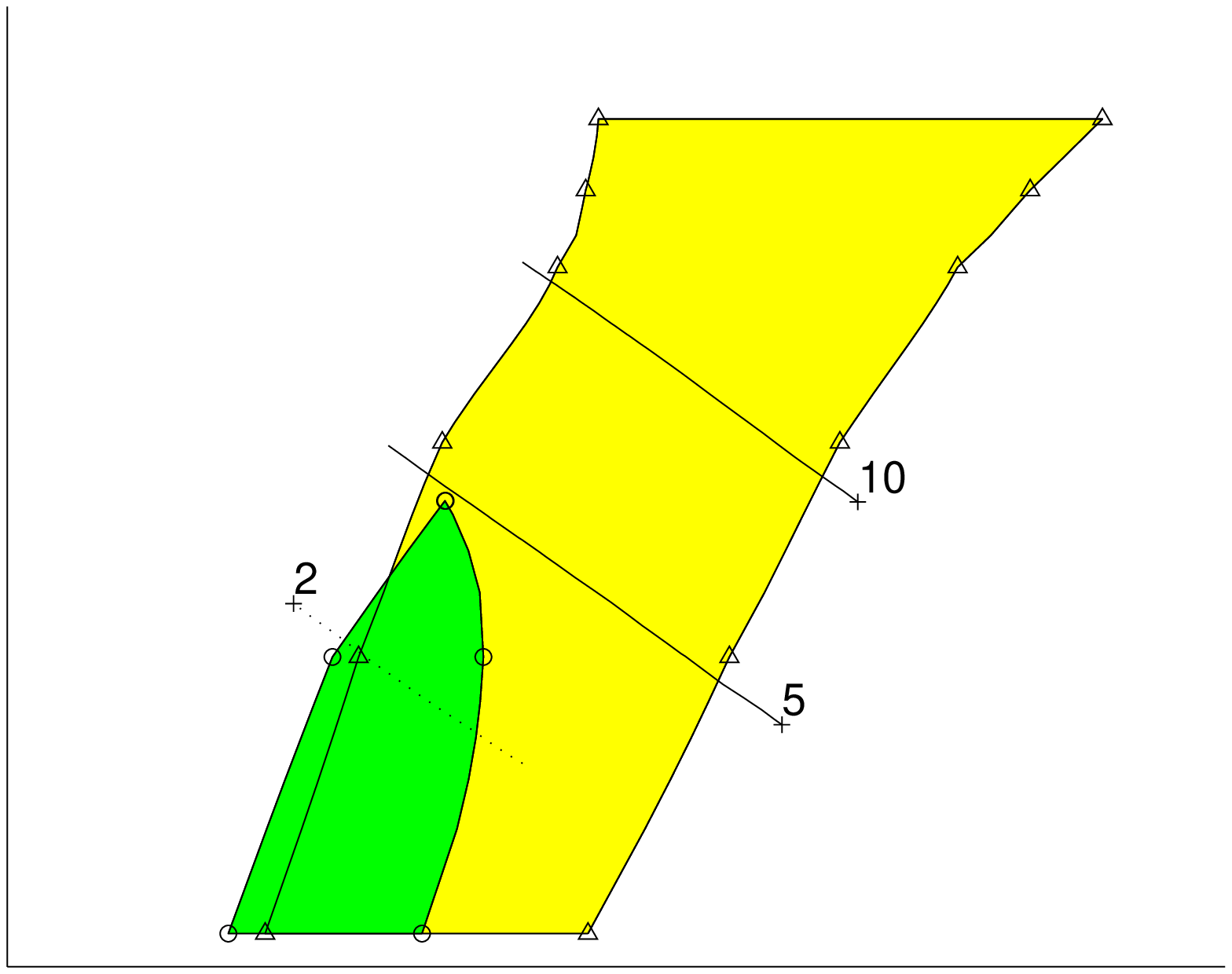}
                       \includegraphics{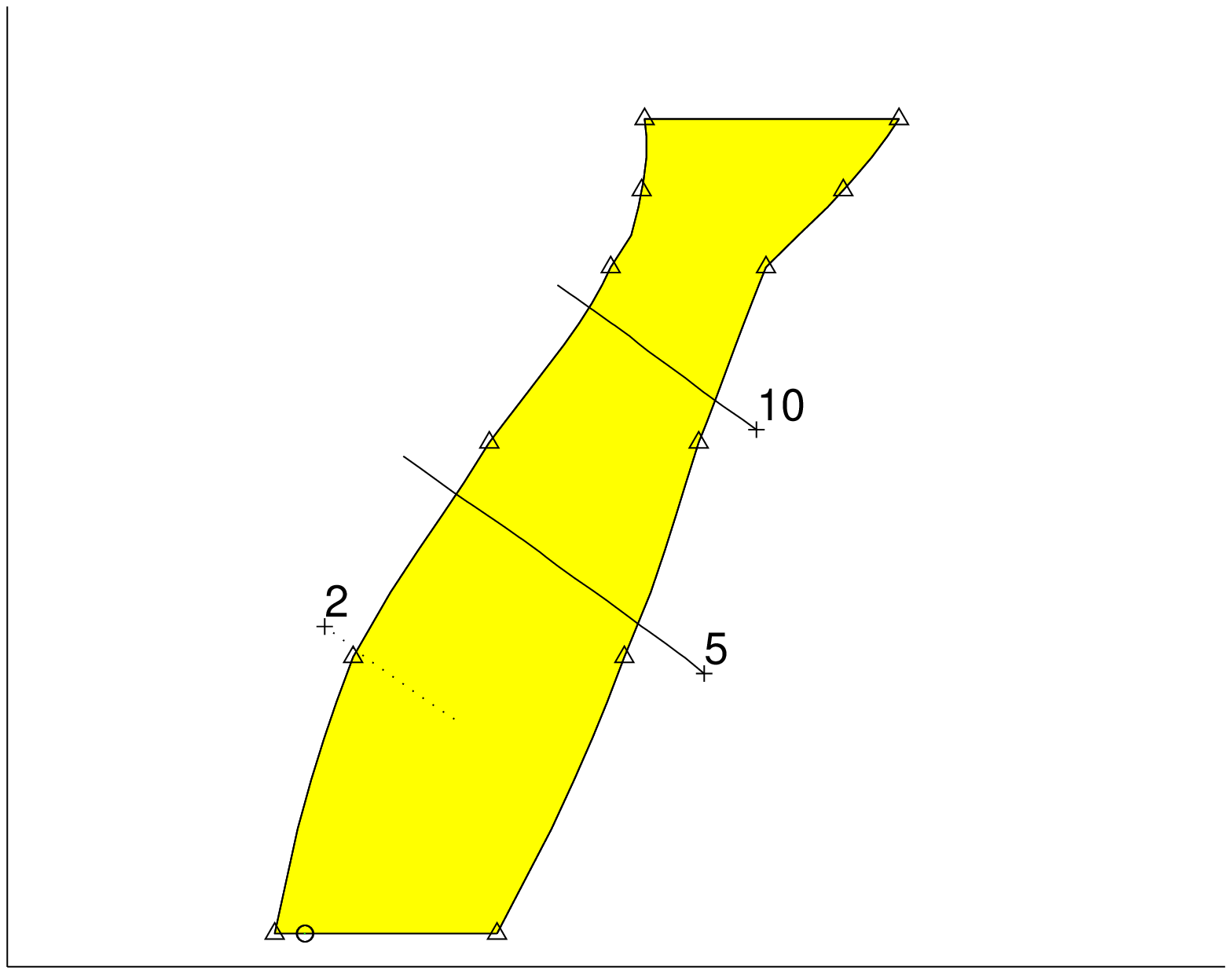}
                       \includegraphics{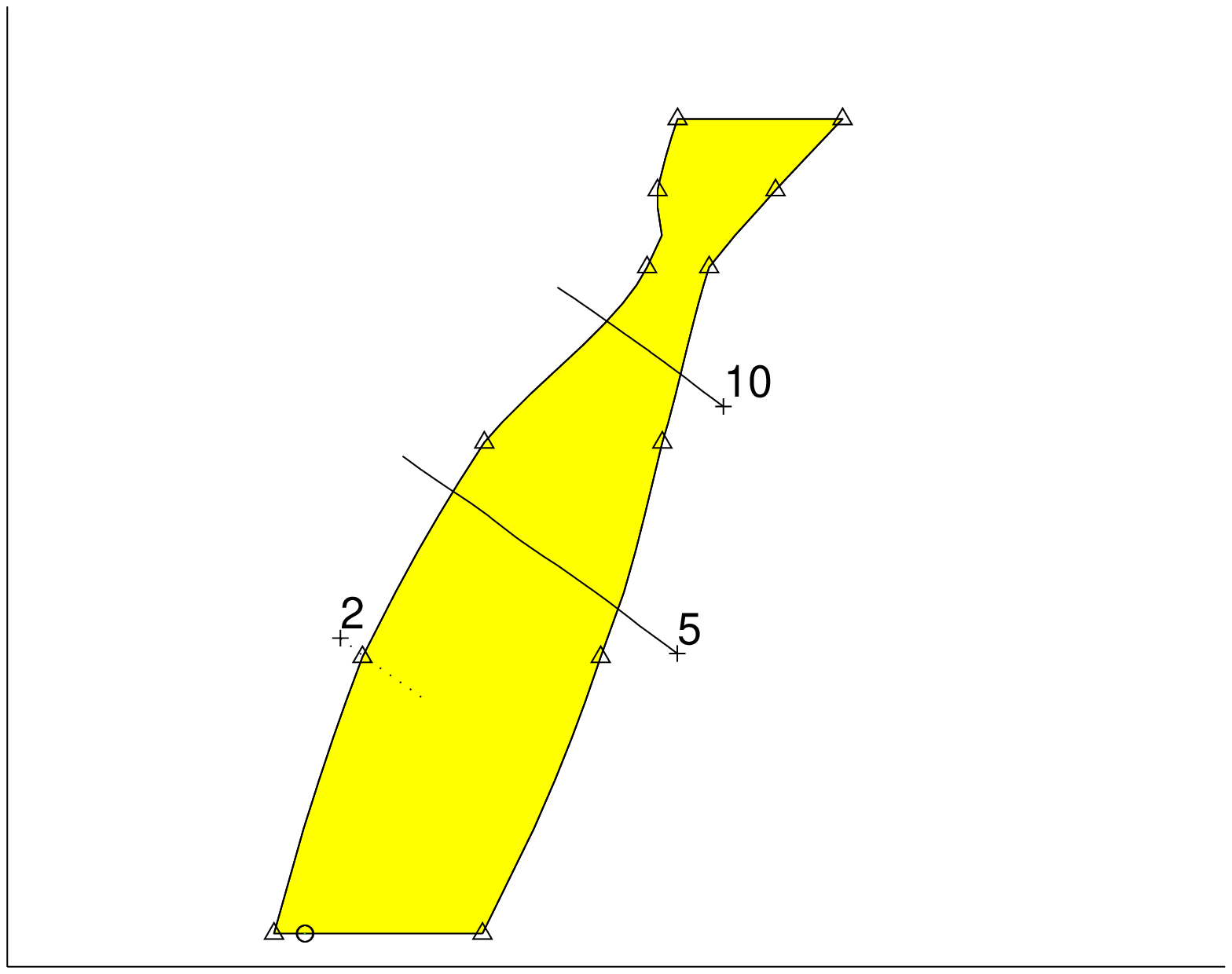}}
 \hfill
 \resizebox{\hsize}{!}{\includegraphics{nullv.ps}
                       \includegraphics{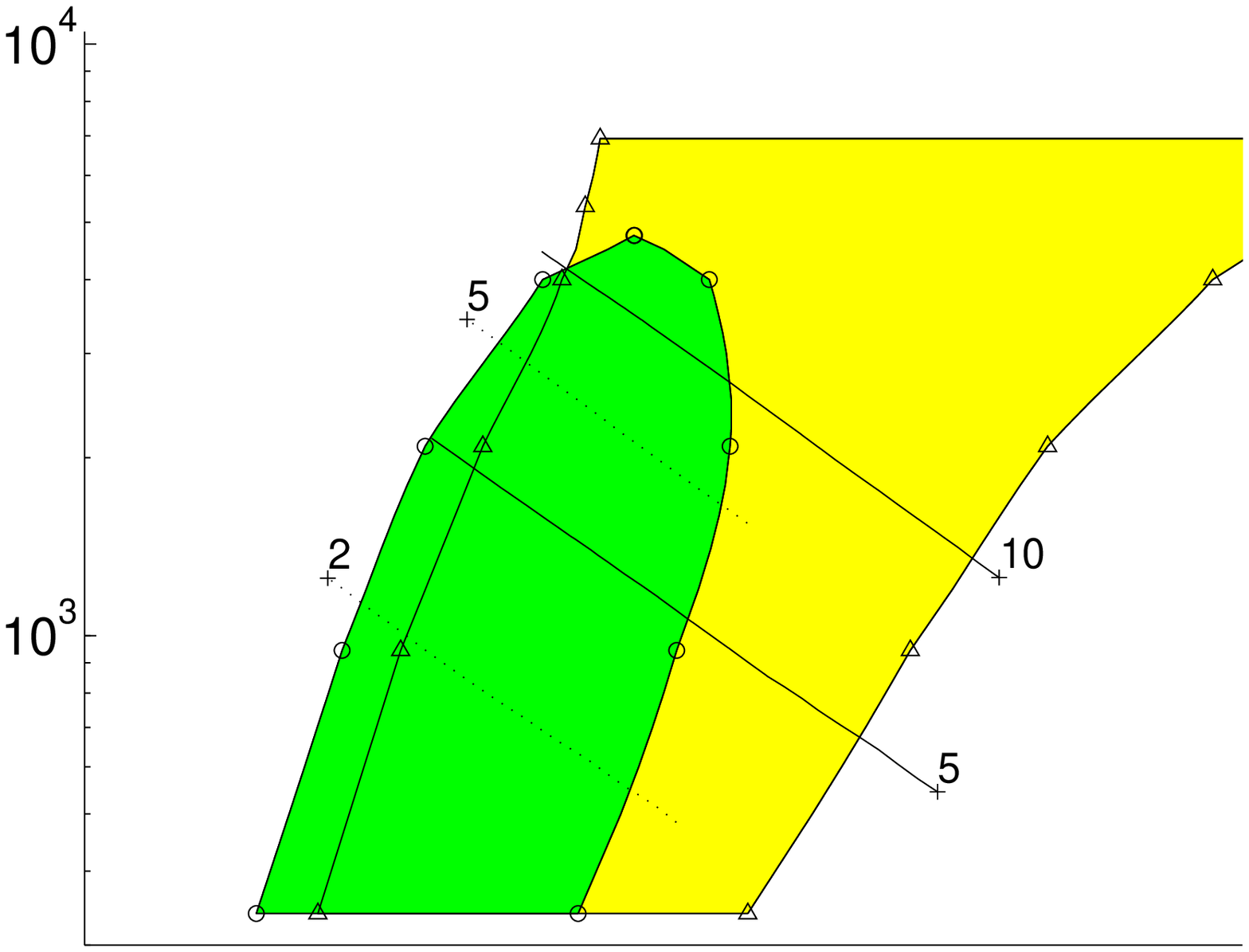}
                       \includegraphics{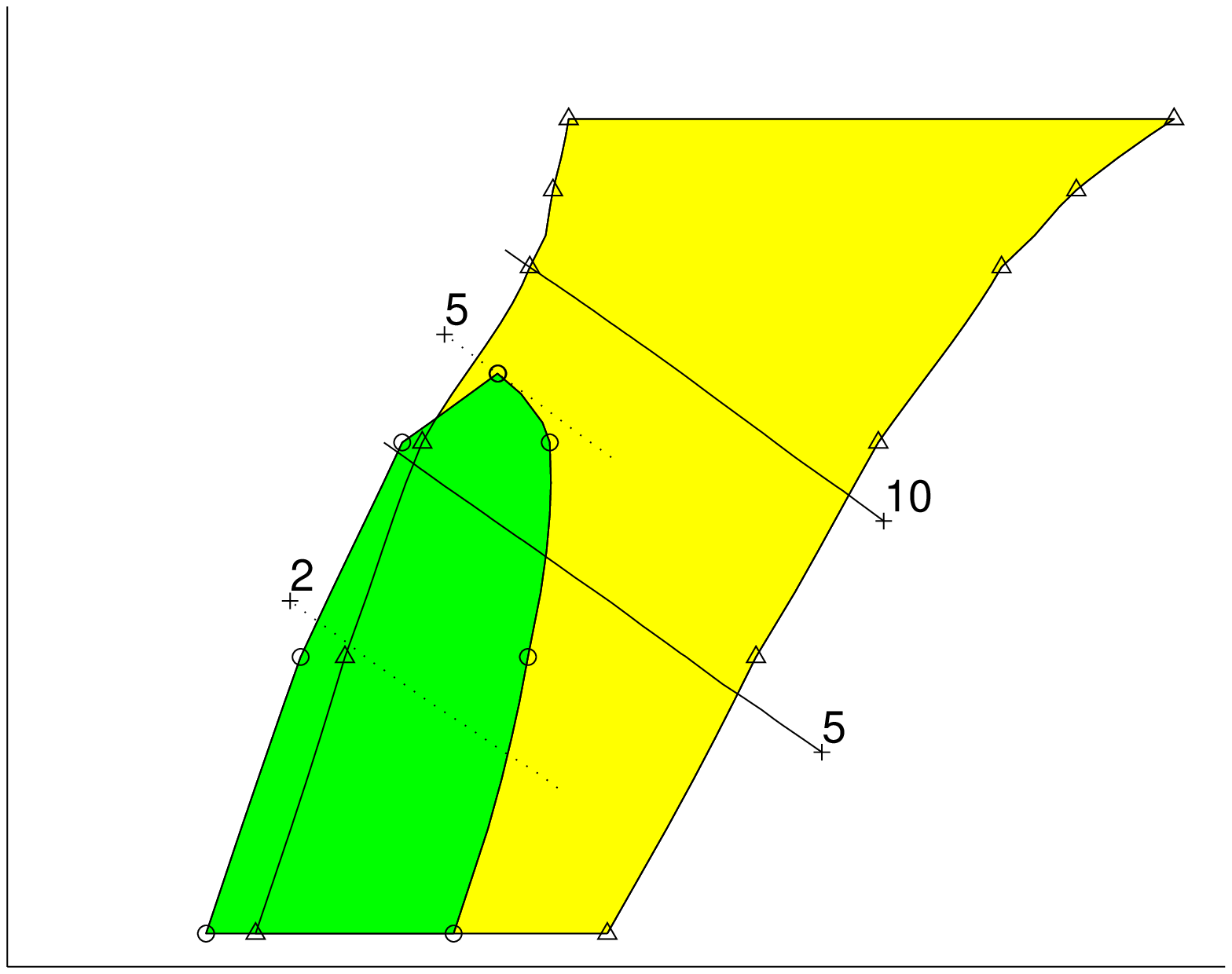}
                       \includegraphics{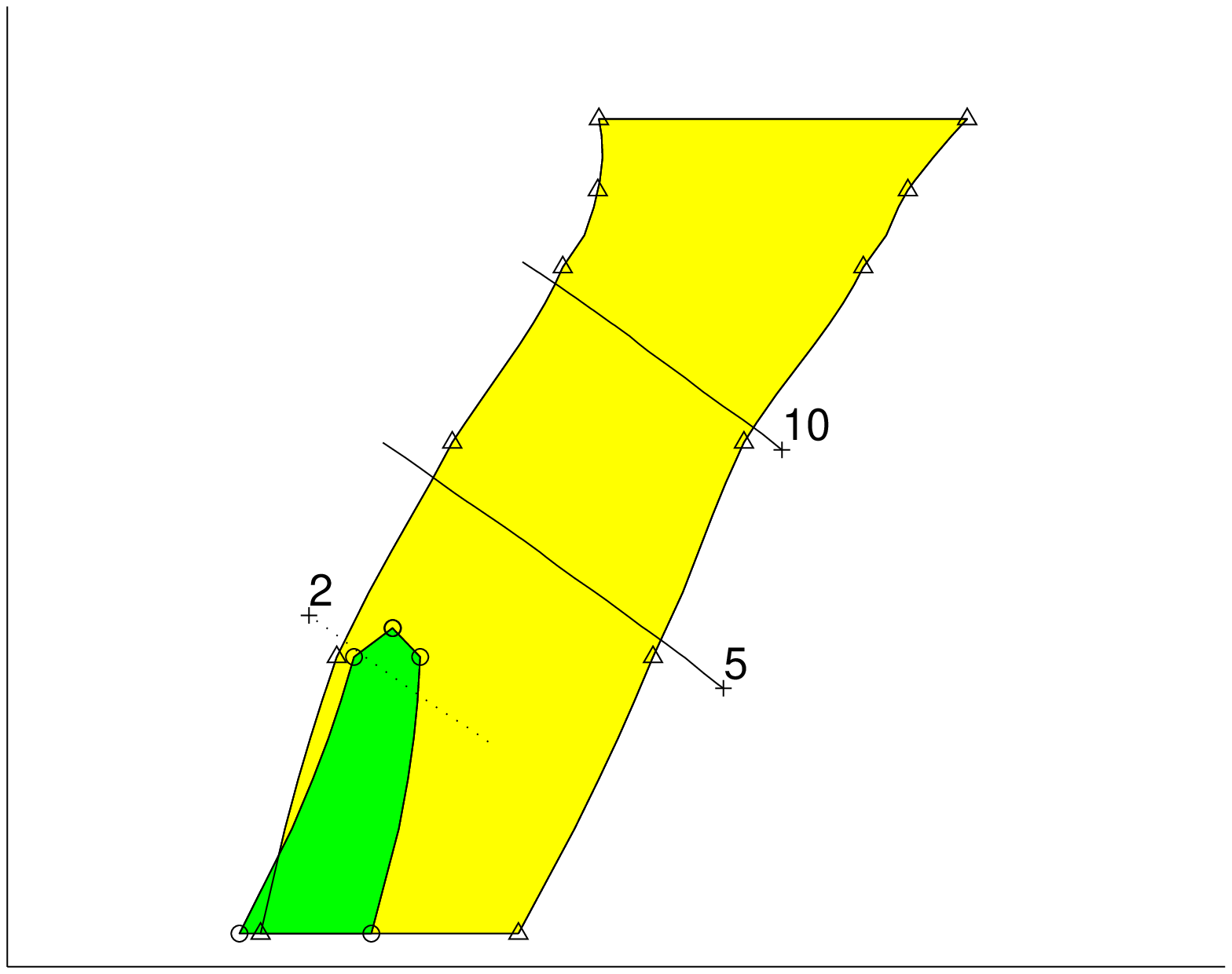}
                       \includegraphics{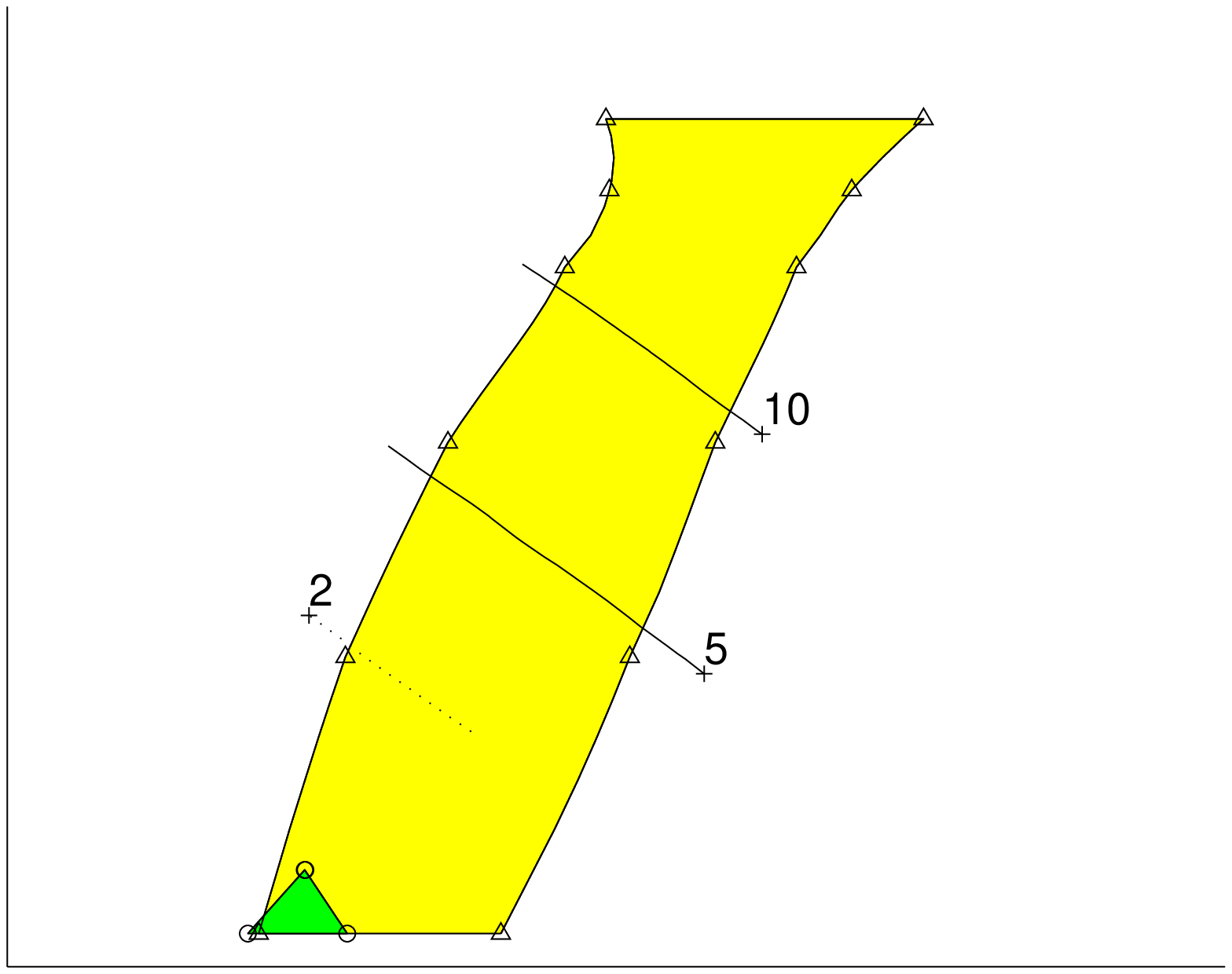}}
 \hfill
 \resizebox{\hsize}{!}{\includegraphics{nullv.ps}
                       \includegraphics{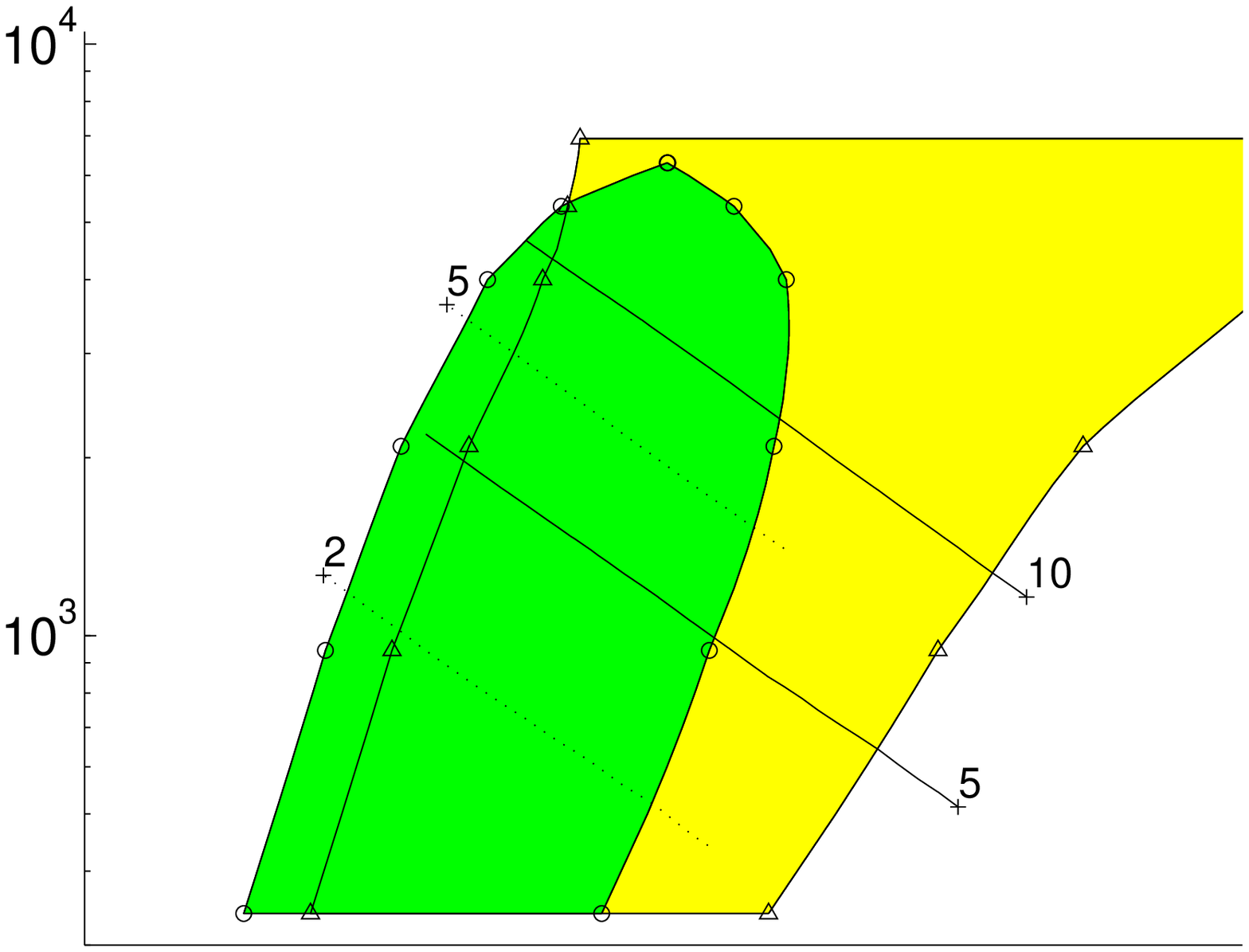}
                       \includegraphics{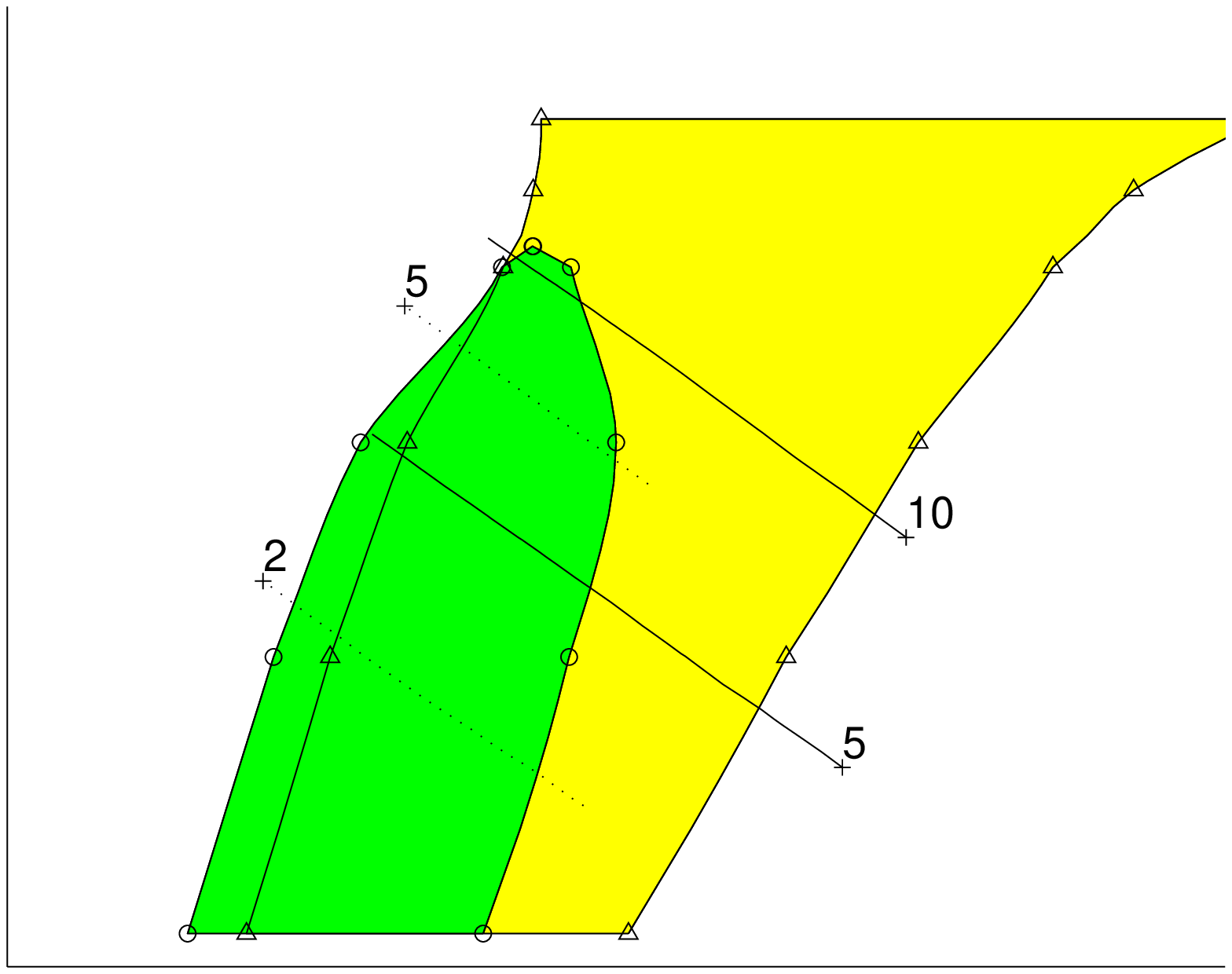}
                       \includegraphics{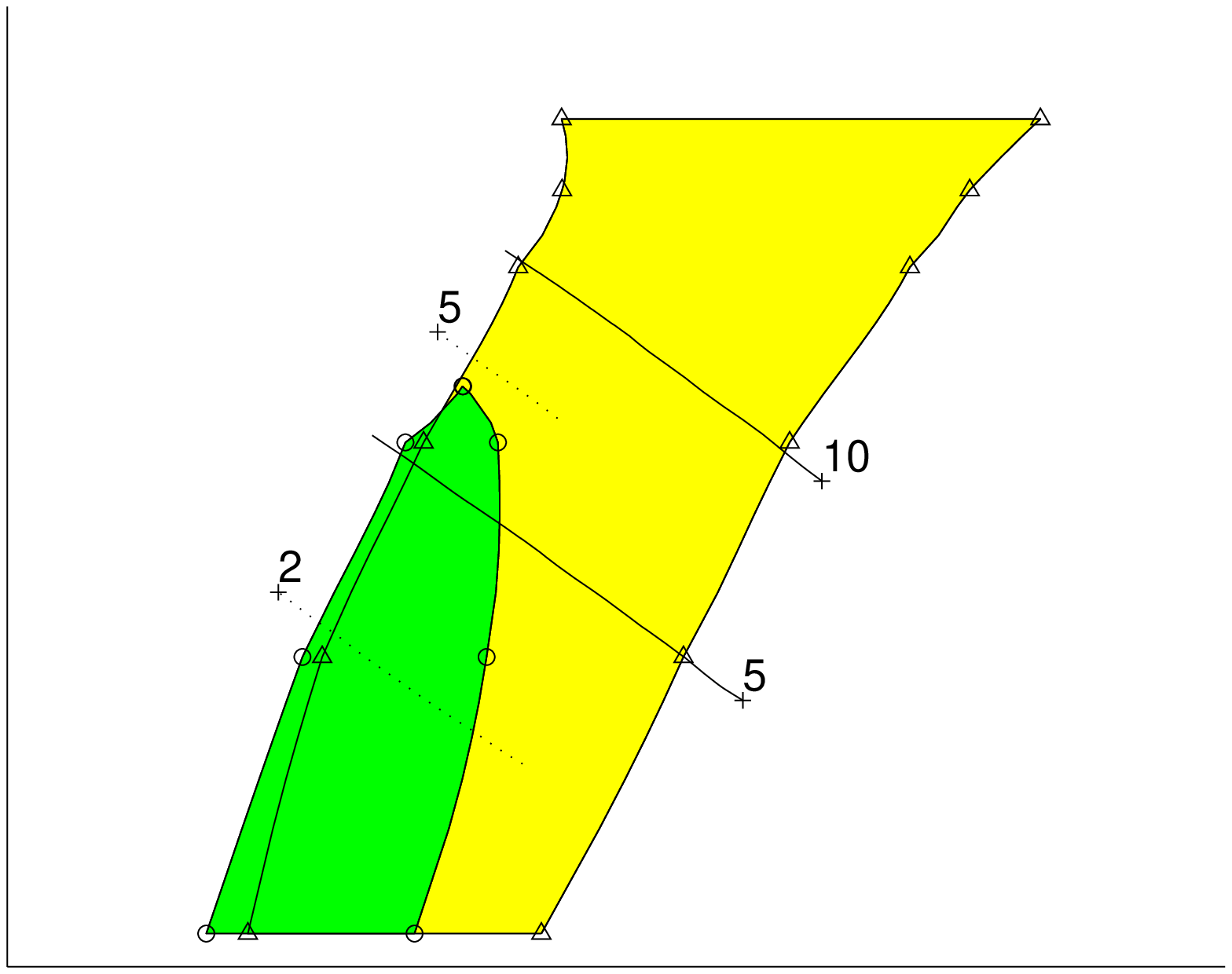}
                       \includegraphics{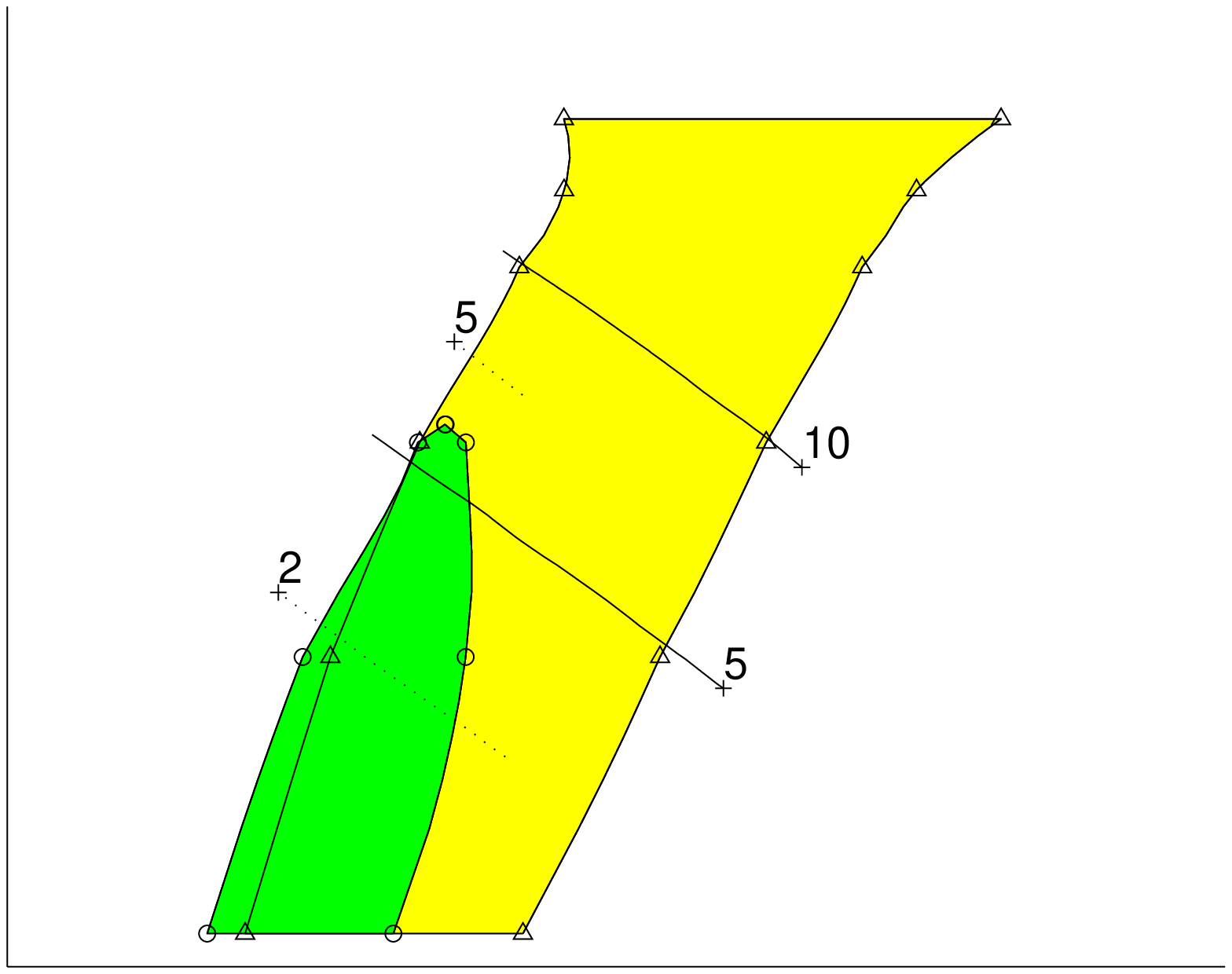}}
 \hfill
 \resizebox{\hsize}{!}{\includegraphics{nullv.ps}
                       \includegraphics{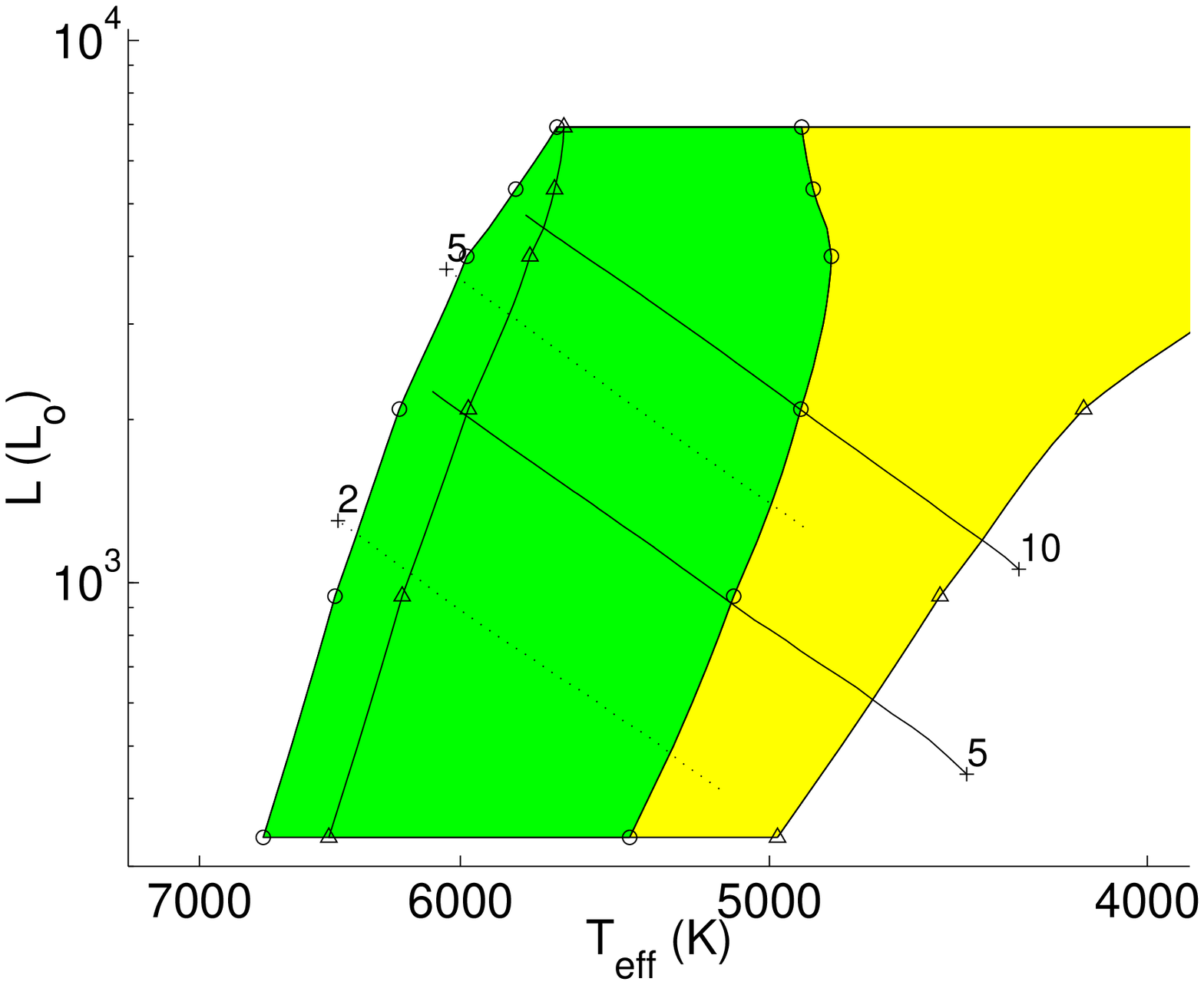}
                       \includegraphics{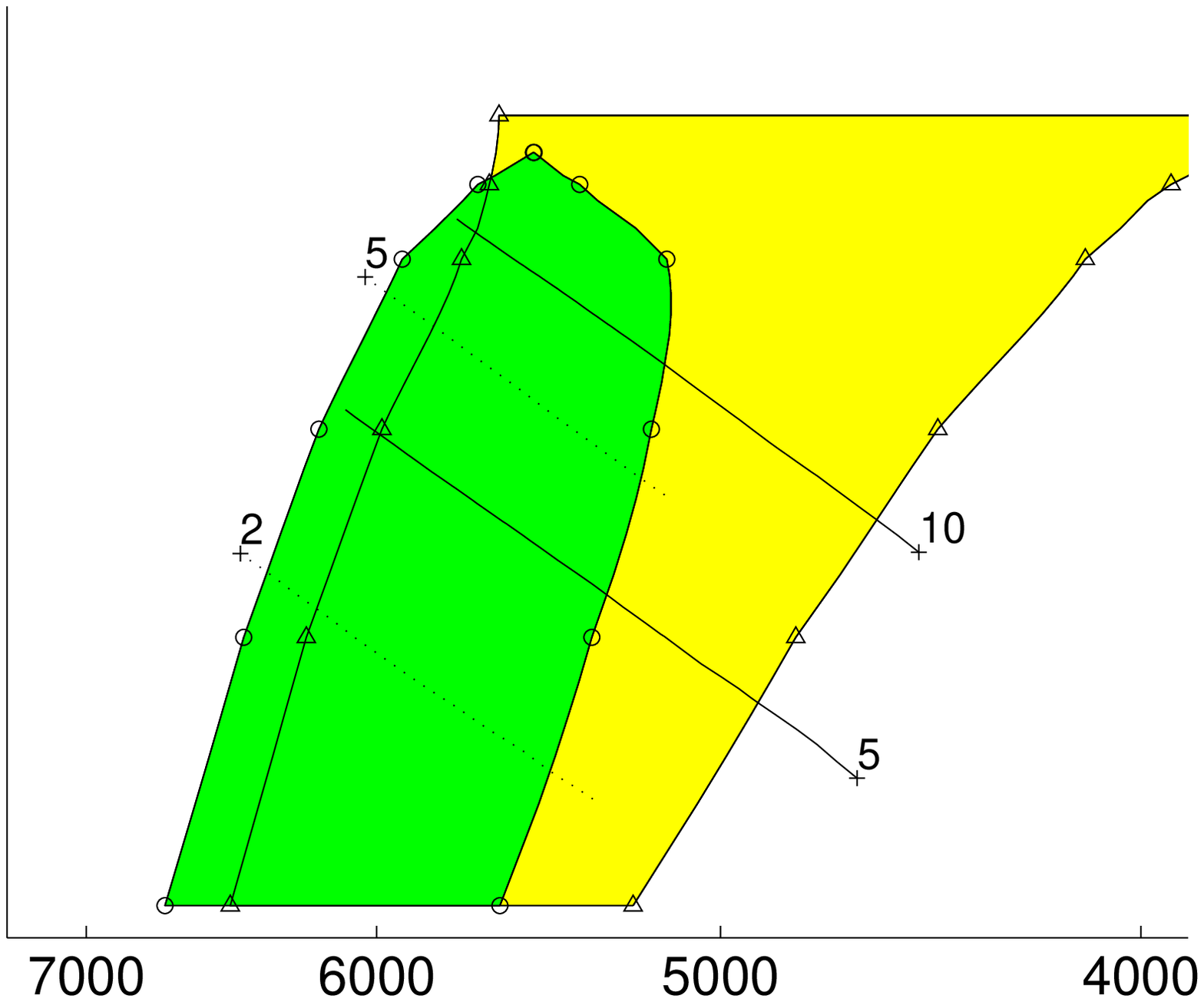}
                       \includegraphics{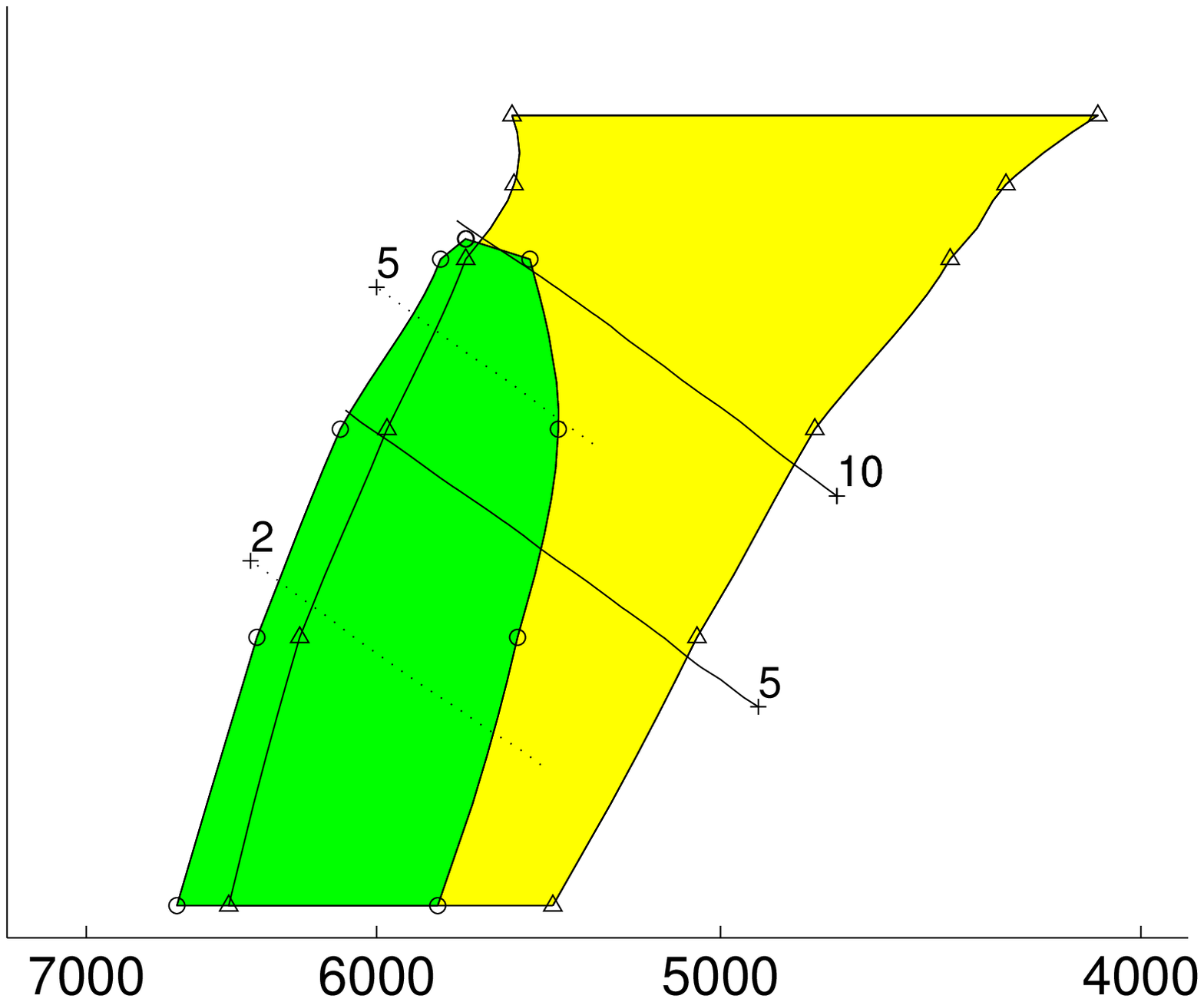}
                       \includegraphics{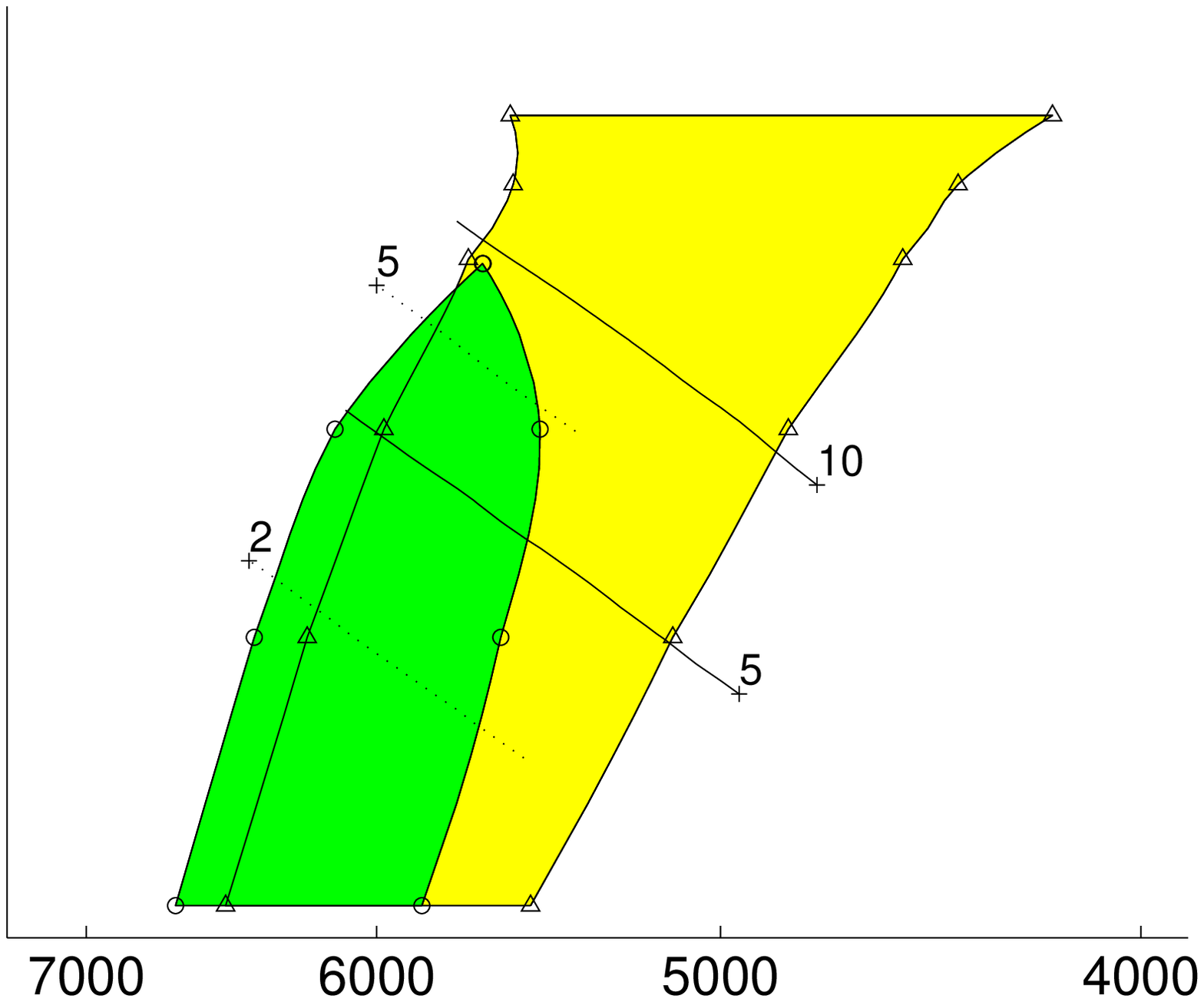}}
 \hfill
 \resizebox{\hsize}{!}{\includegraphics{nullhv.ps}
                       \includegraphics{nullh.ps}
                       \includegraphics{nullh.ps}
                       \includegraphics{nullh.ps}
                       \includegraphics{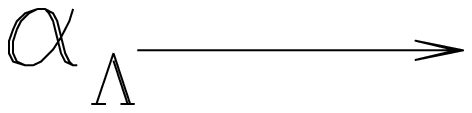}
                       }
 \vskip 0pt
 \parbox[b]{\hsize}{
   \caption{
 Linear instability strips as a function of $\alpha_\nu$  and
 $\alpha_\Lambda$ with fixed $\alpha_c$ = 3.0,
   fundamental IS (light shaded area) and first-overtone IS (dark area).
 {\sl Bottom to top:} $\alpha_\nu$ = 1.5, 2.0, 2.5, 3.0;  
 {\sl left to right:} $\alpha_\Lambda$ = 0.281, 0.328, 0.375, 0.390.
   }
 \label{mltis}
 }
 \end{figure*}
 % END FIGURE 13

 % BEGIN FIGURE 14
 \begin{figure*}
 \resizebox{\hsize}{!}{\includegraphics{alphan.ps}
                       \includegraphics{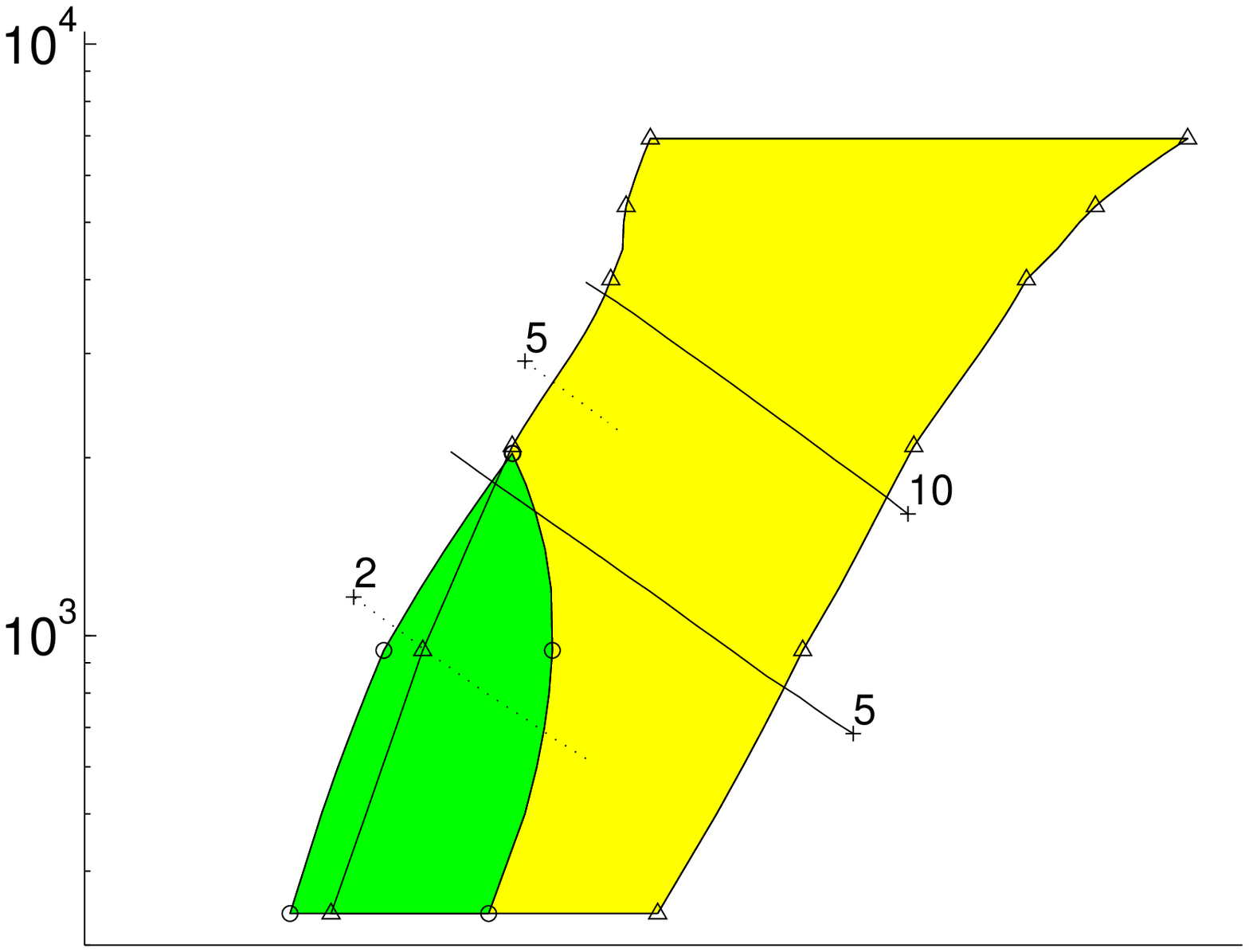}
                       \includegraphics{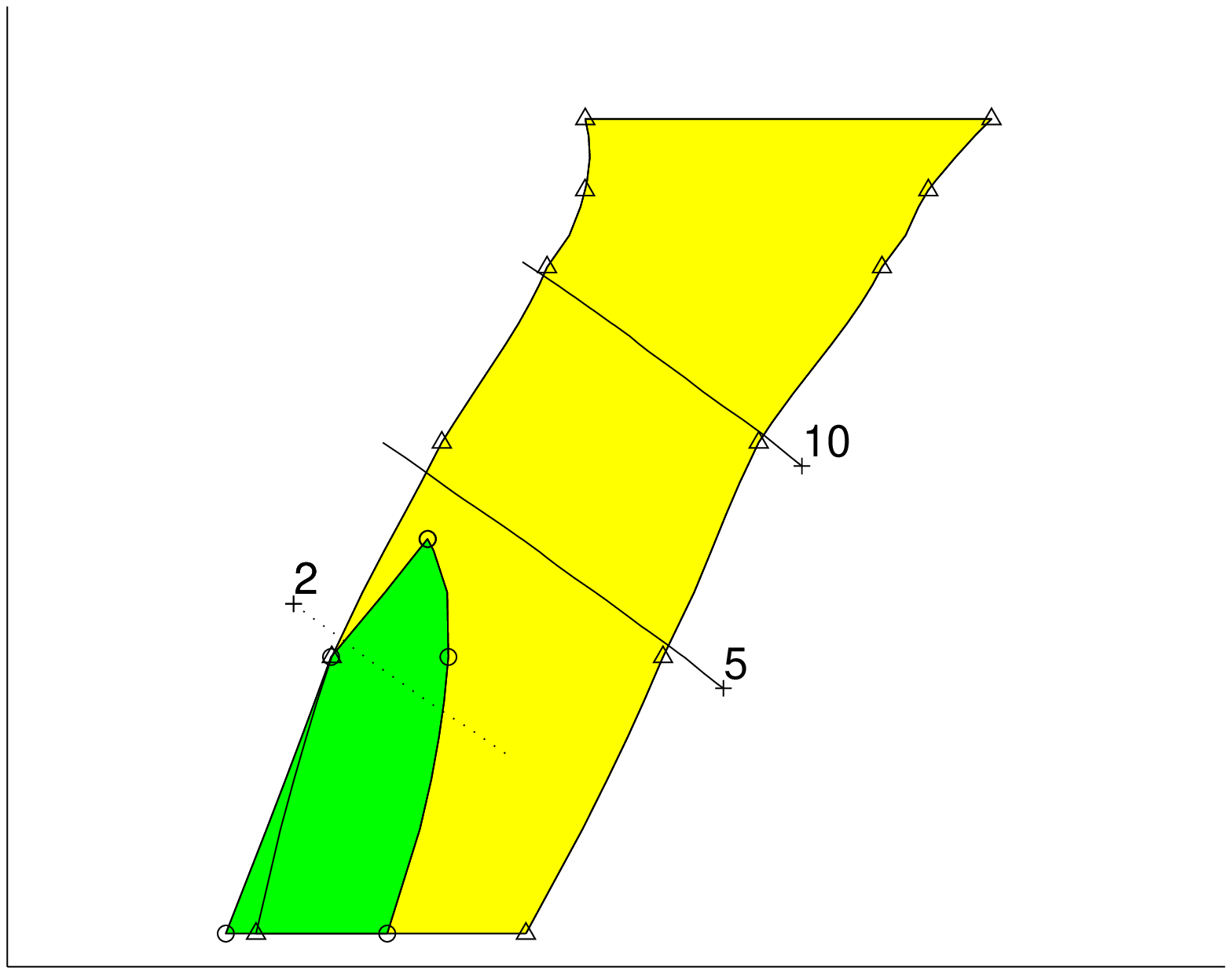}
                       \includegraphics{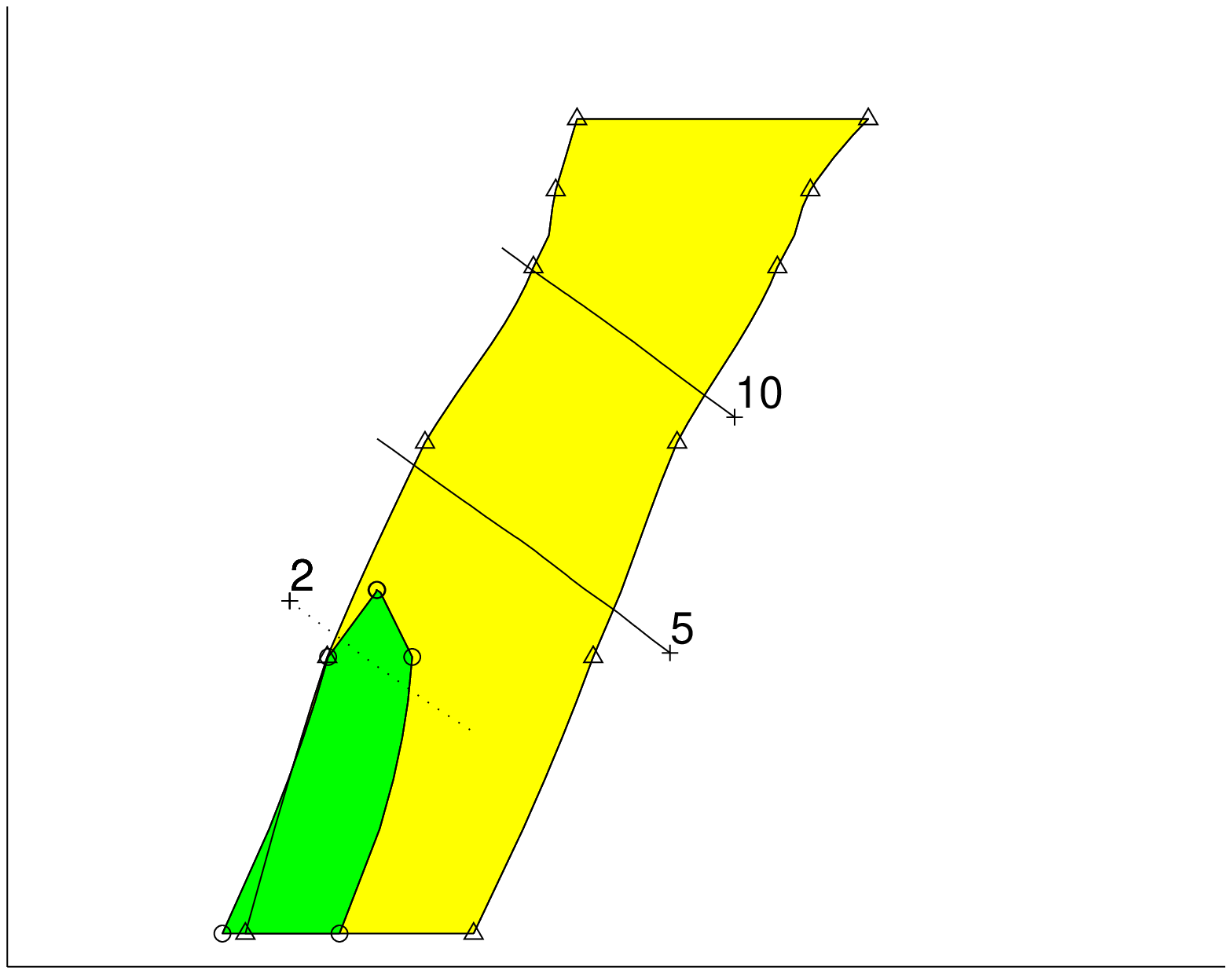}
                       \includegraphics{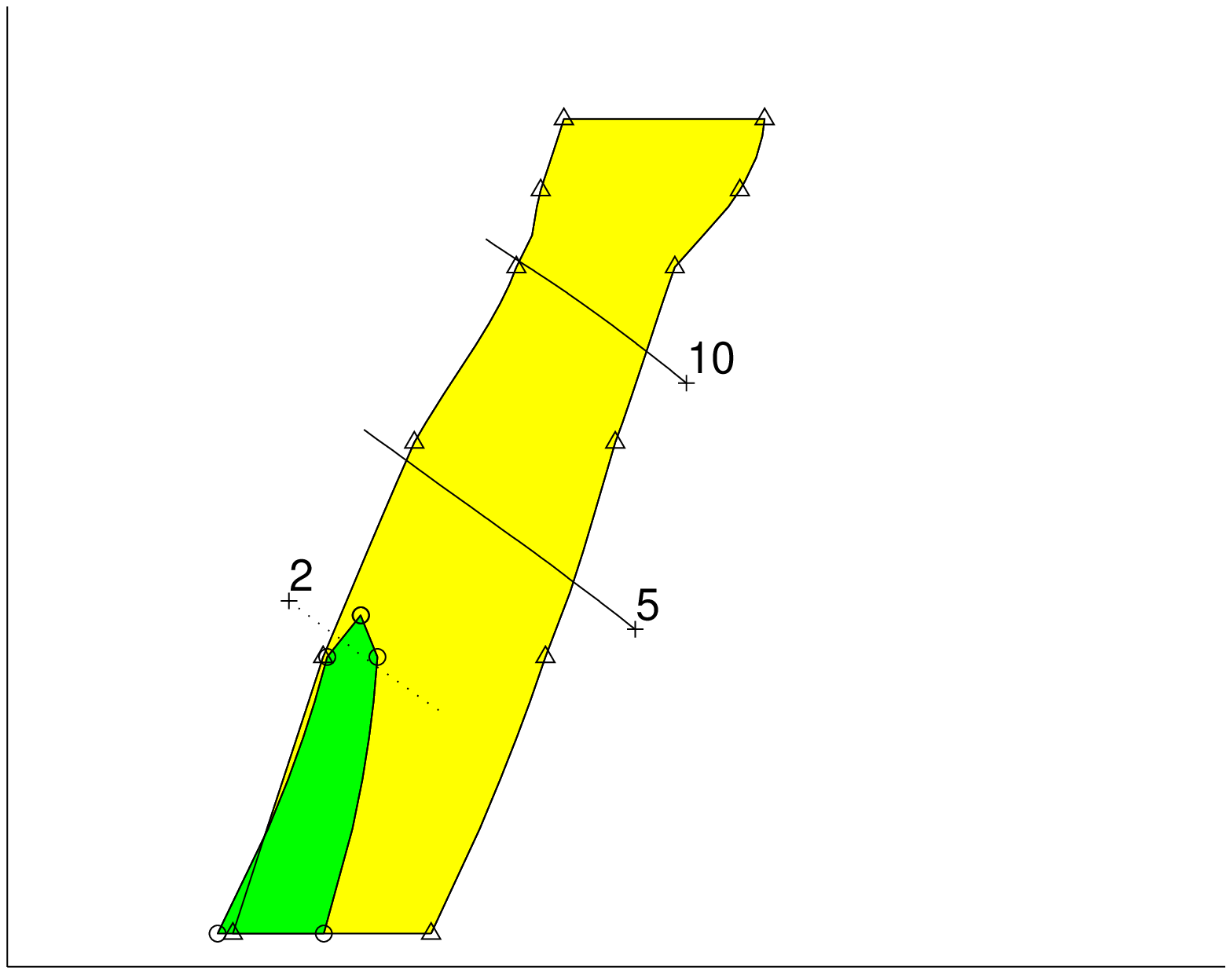}}
 \hfill
 \resizebox{\hsize}{!}{\includegraphics{nullv.ps}
                       \includegraphics{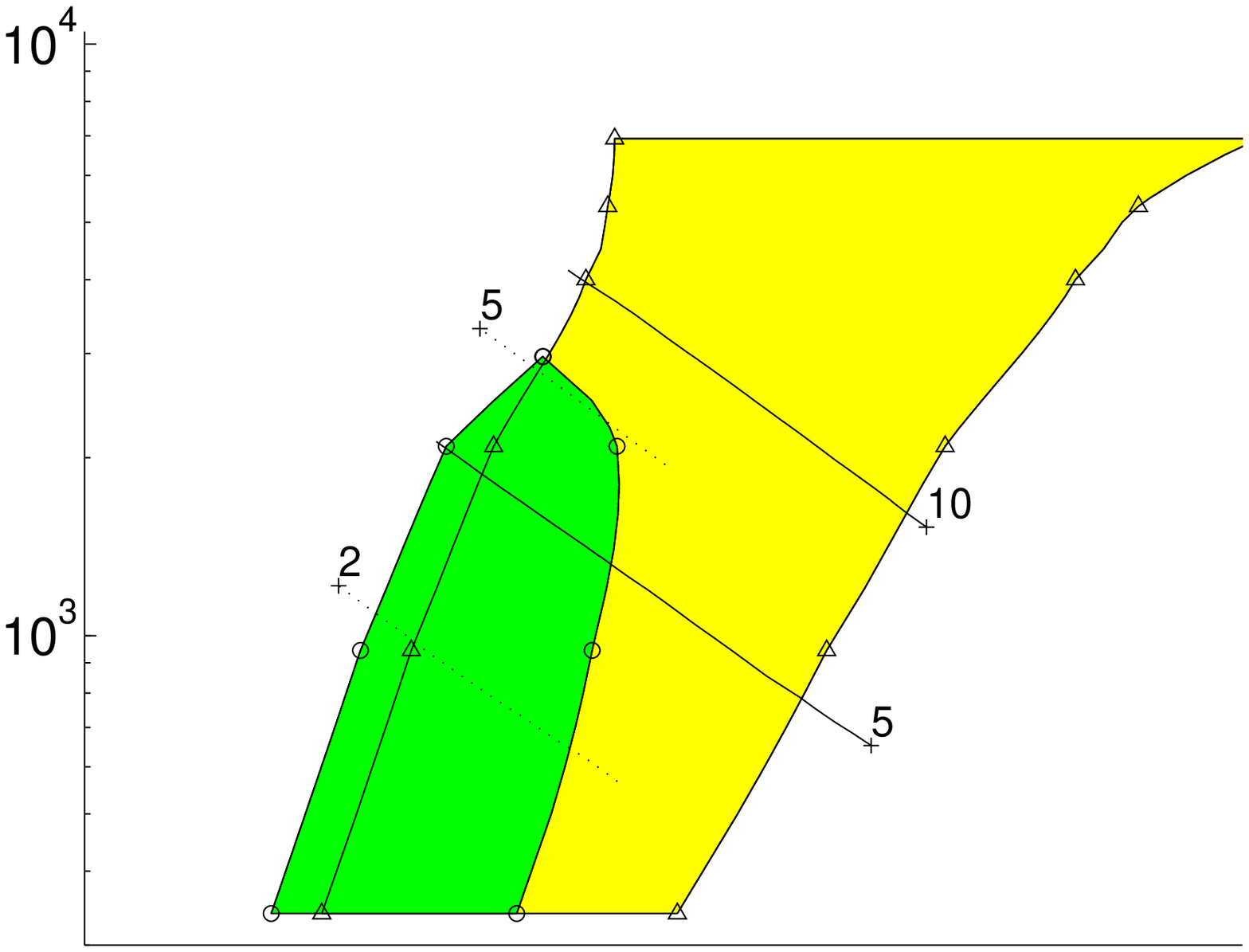}
                       \includegraphics{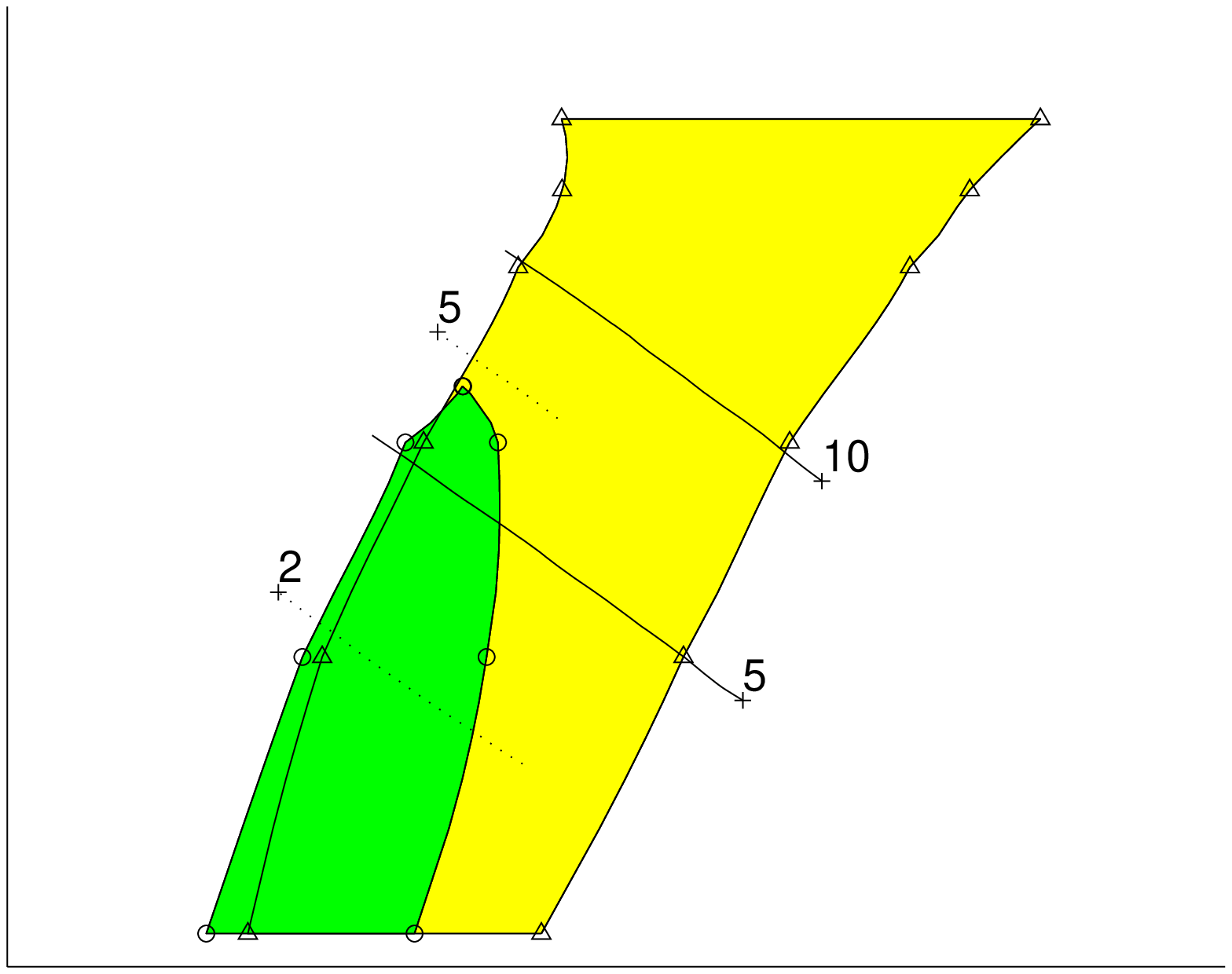}
                       \includegraphics{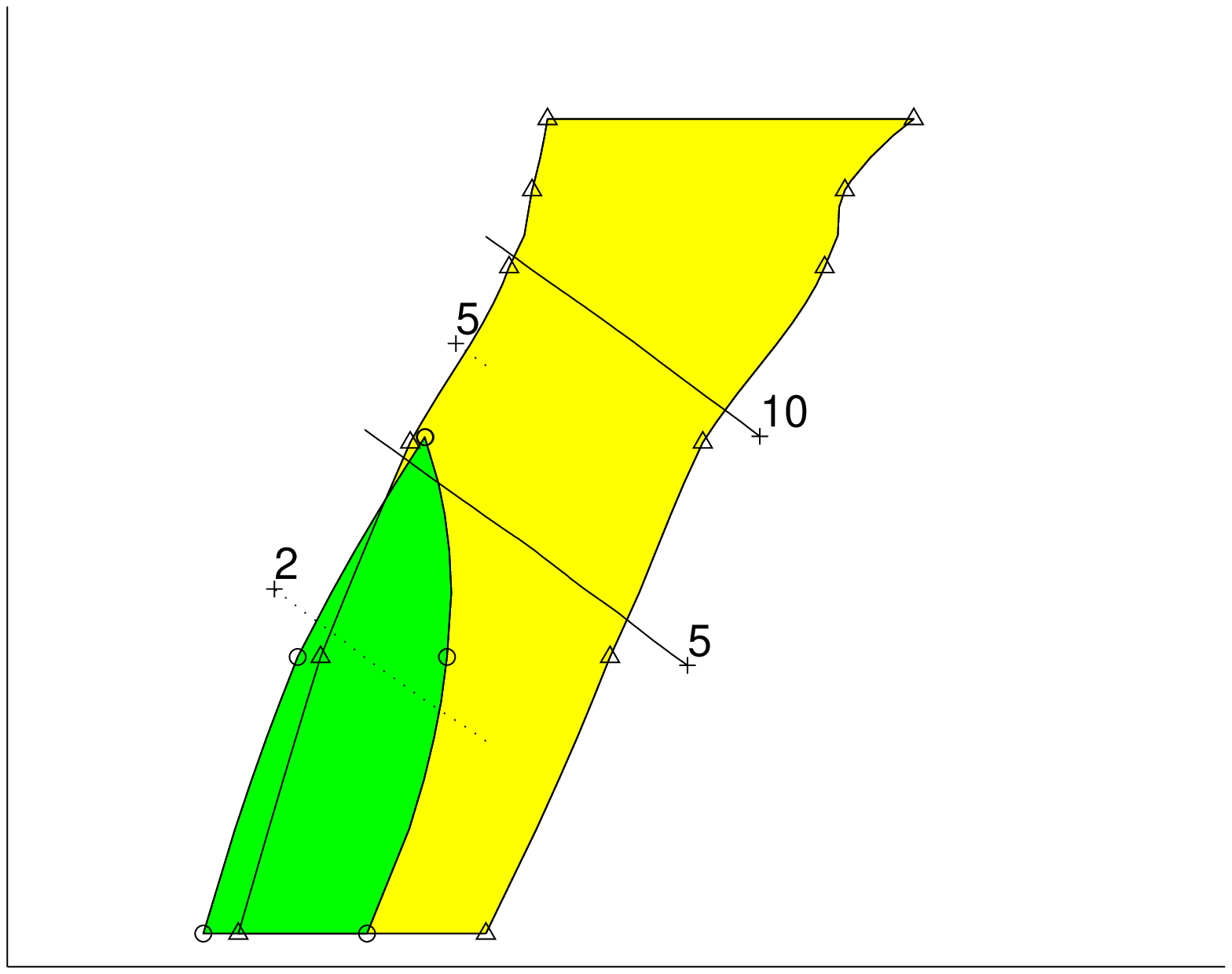}
                       \includegraphics{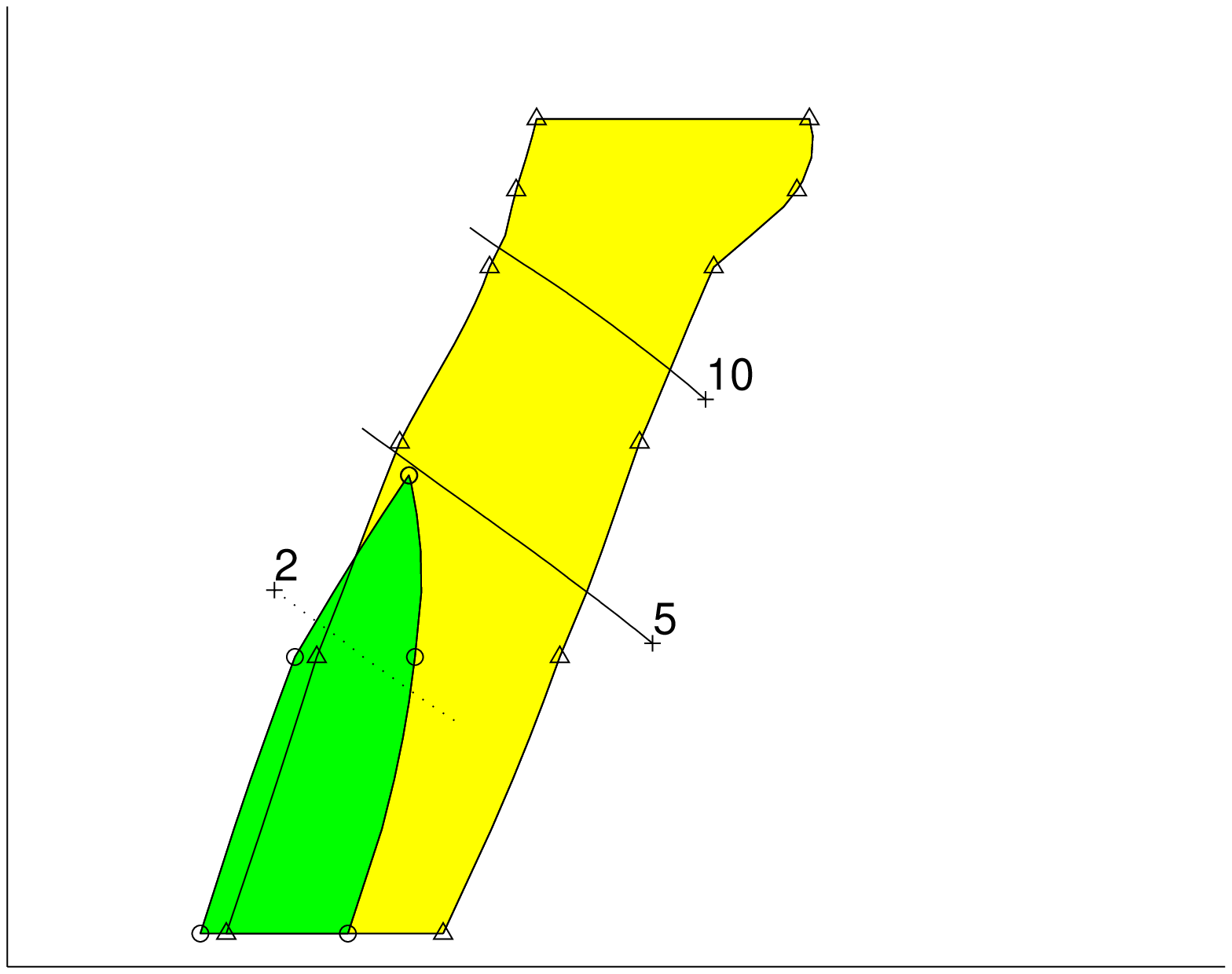}}
 \hfill
 \resizebox{\hsize}{!}{\includegraphics{nullv.ps}
                       \includegraphics{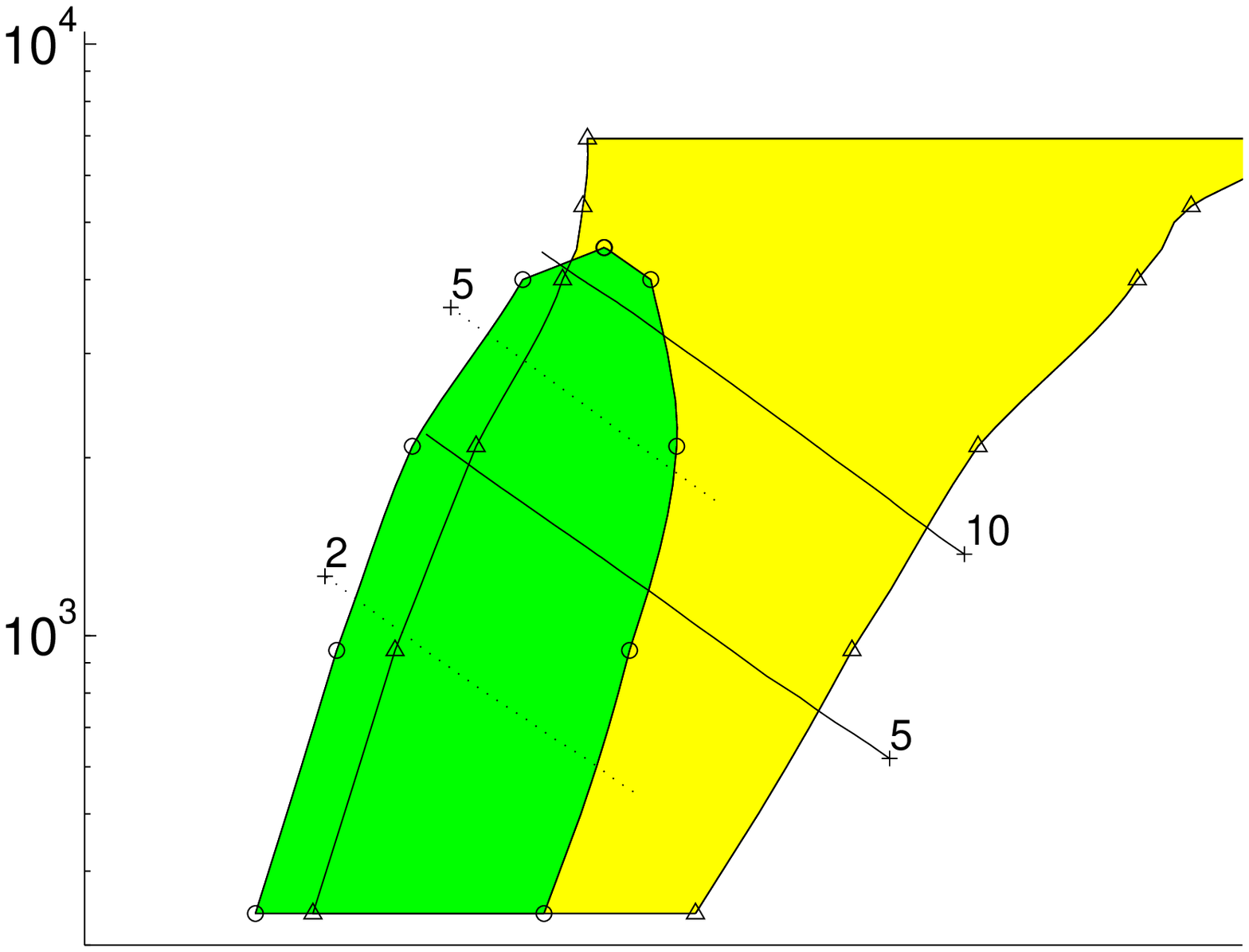}
                       \includegraphics{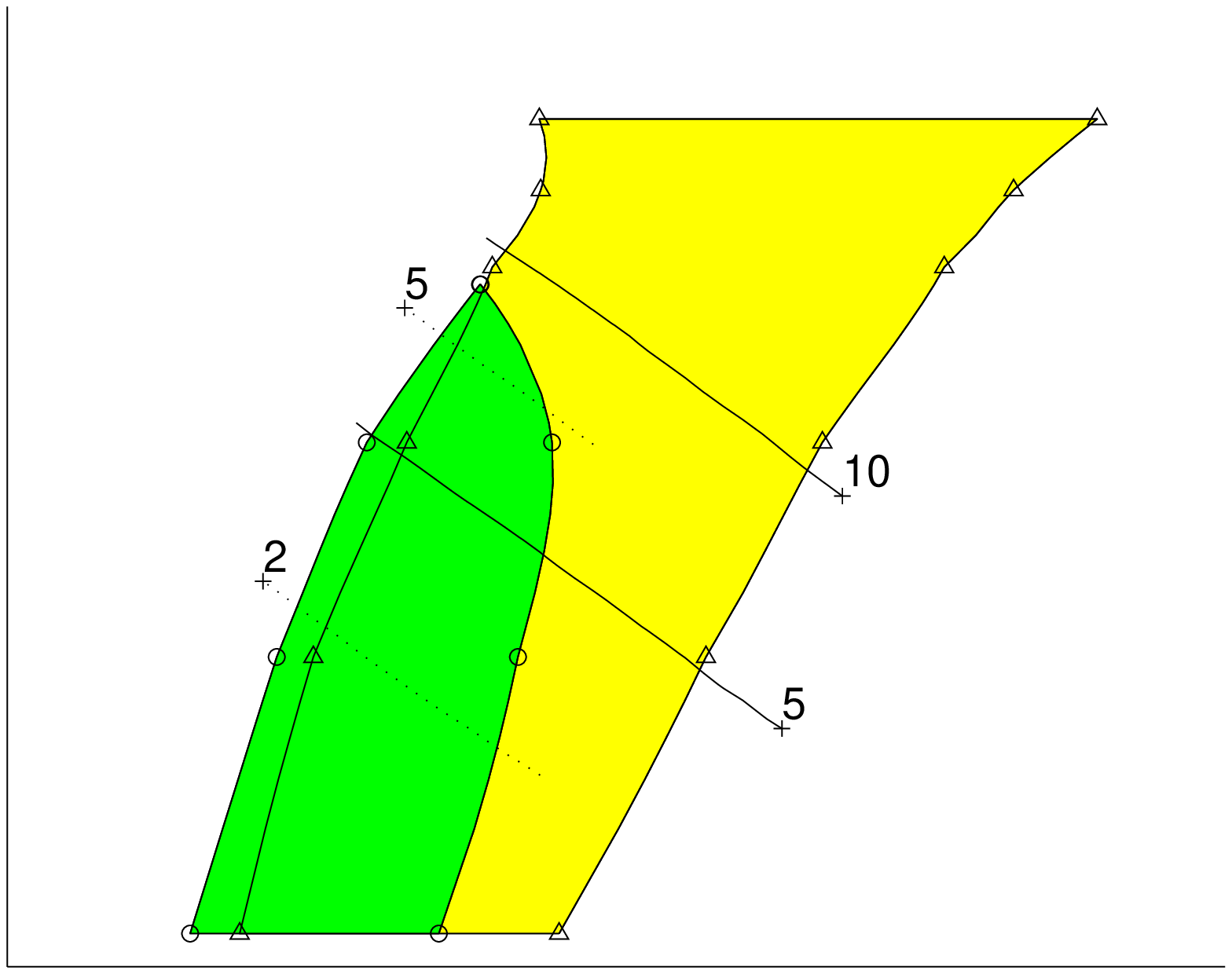}
                       \includegraphics{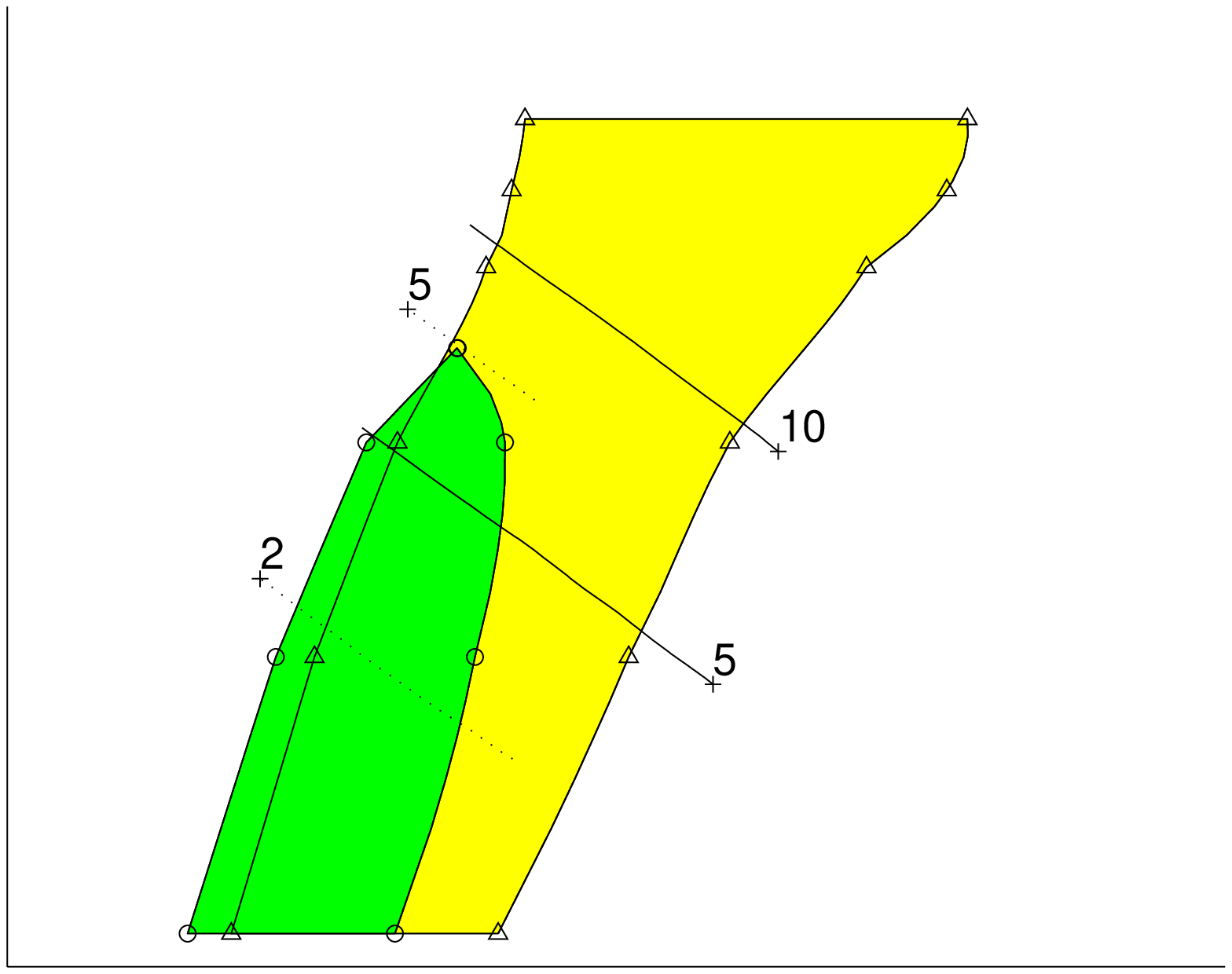}
                       \includegraphics{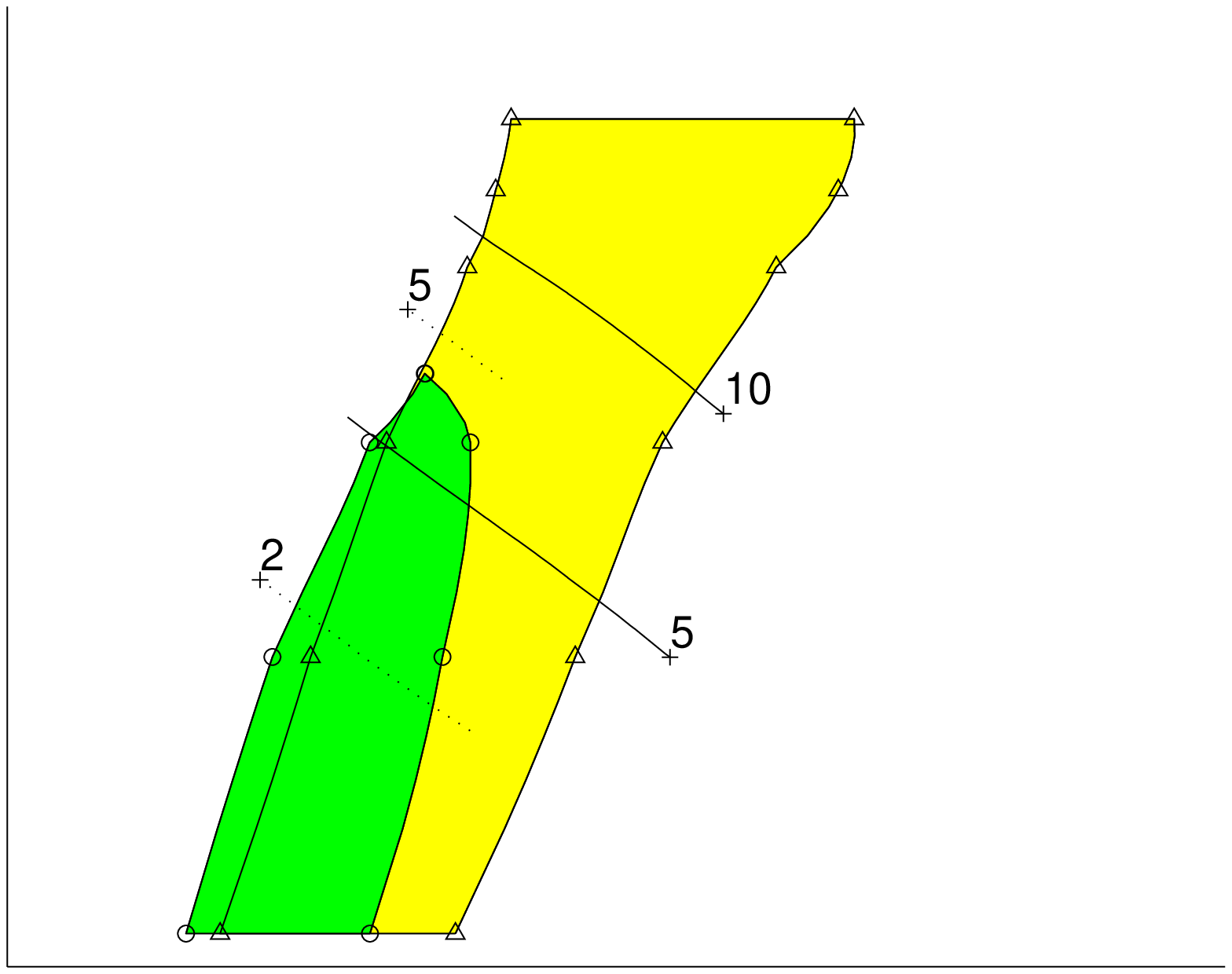}}
 \hfill
 \resizebox{\hsize}{!}{\includegraphics{nullv.ps}
                       \includegraphics{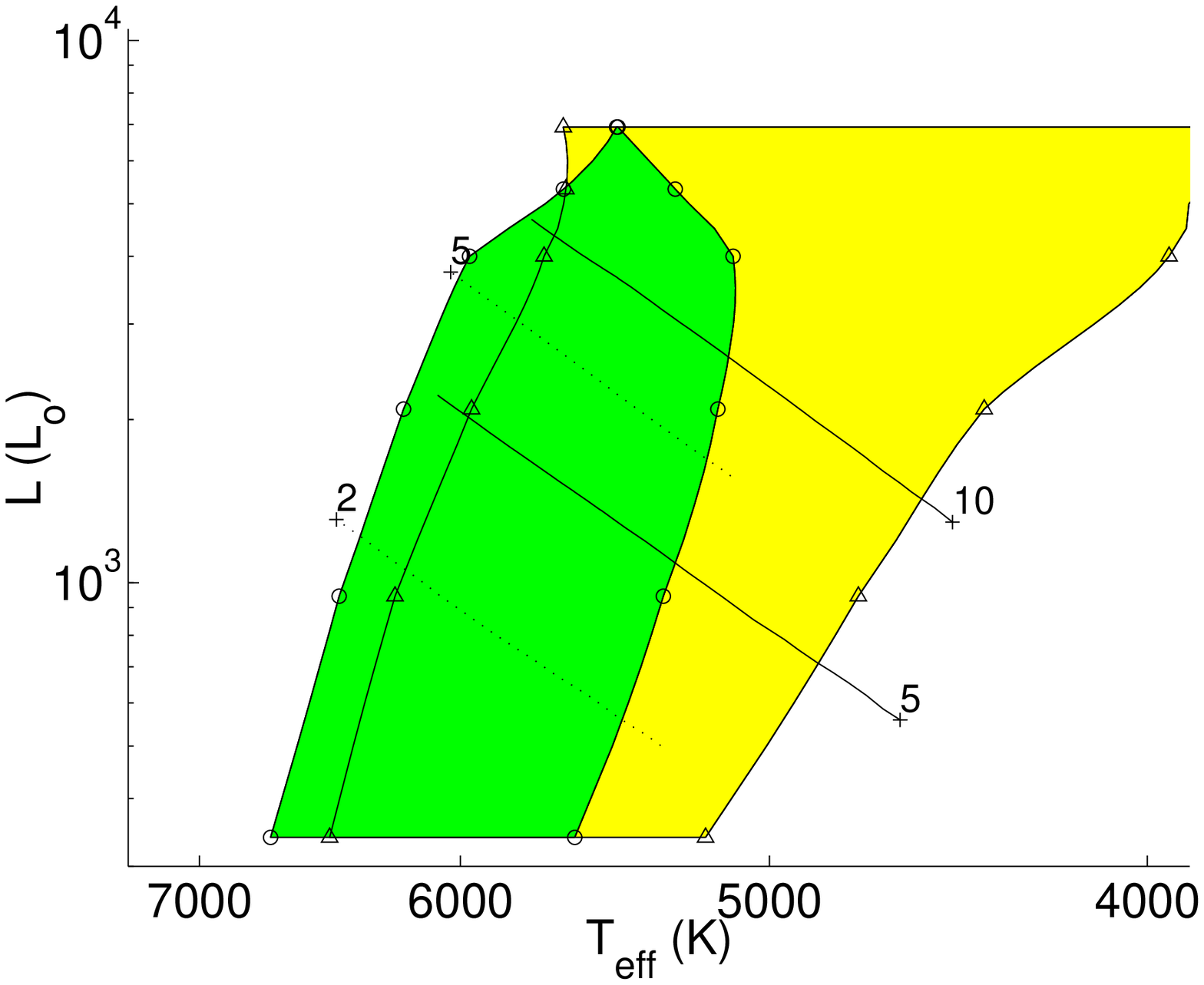}
                       \includegraphics{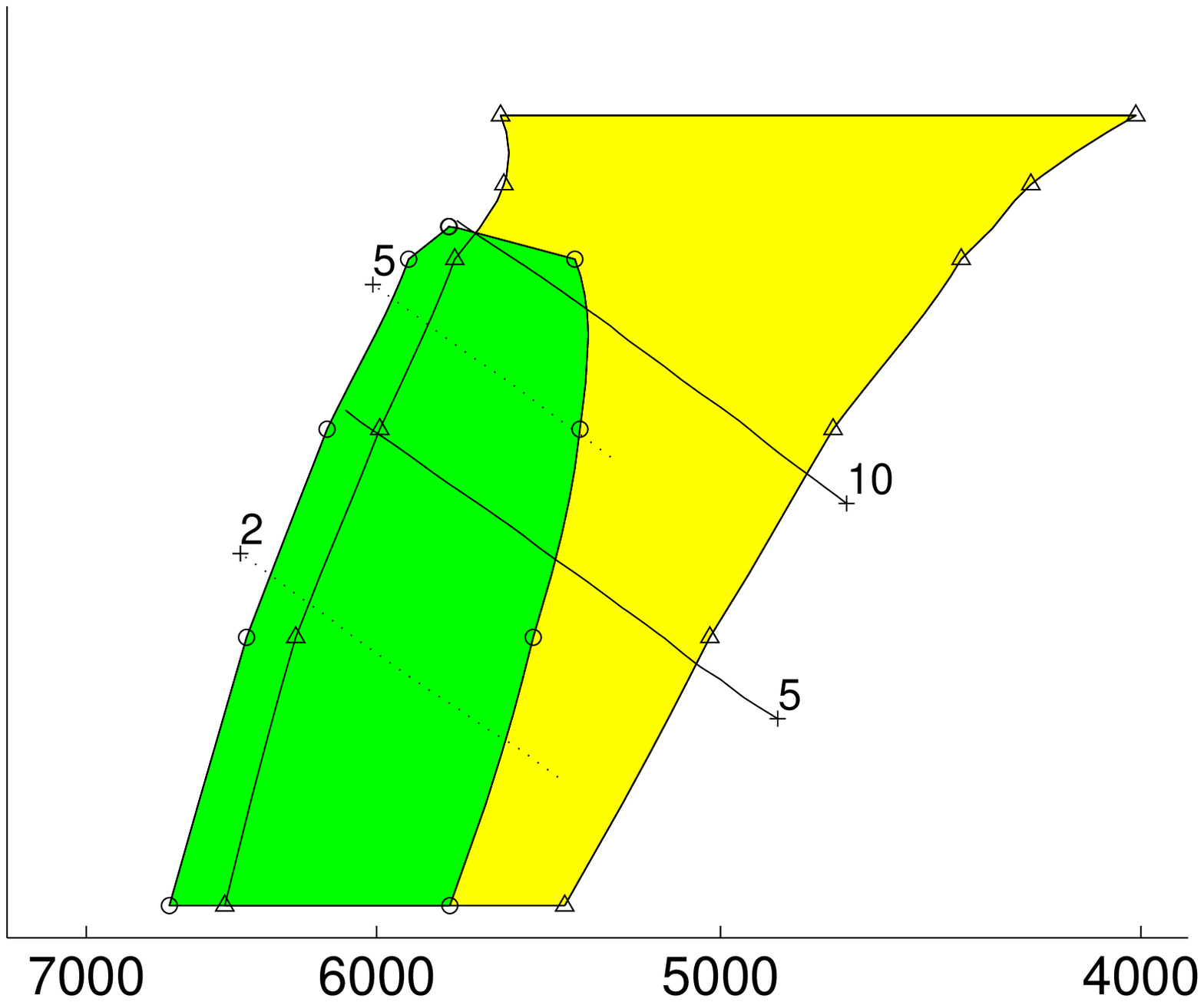}
                       \includegraphics{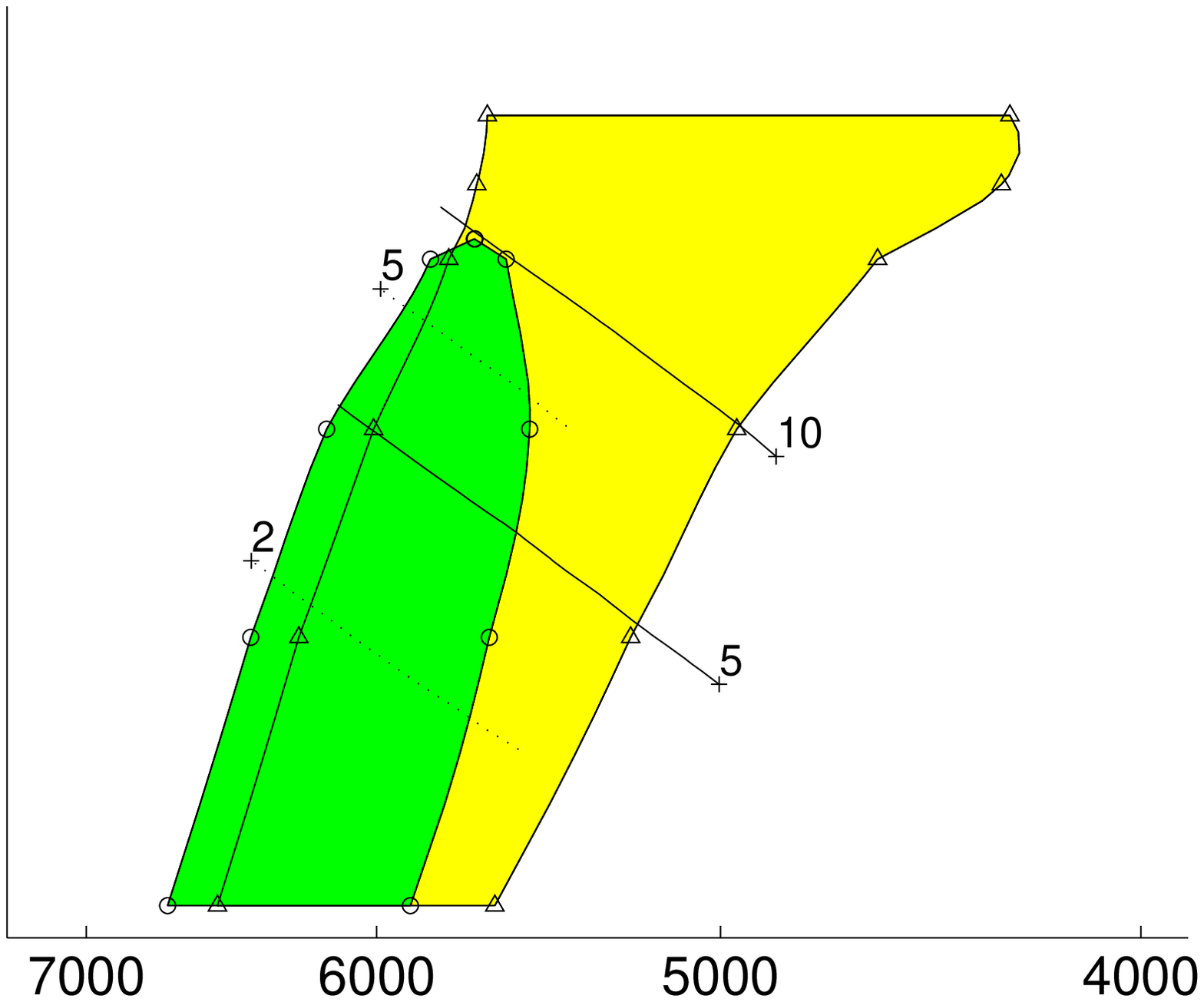}
                       \includegraphics{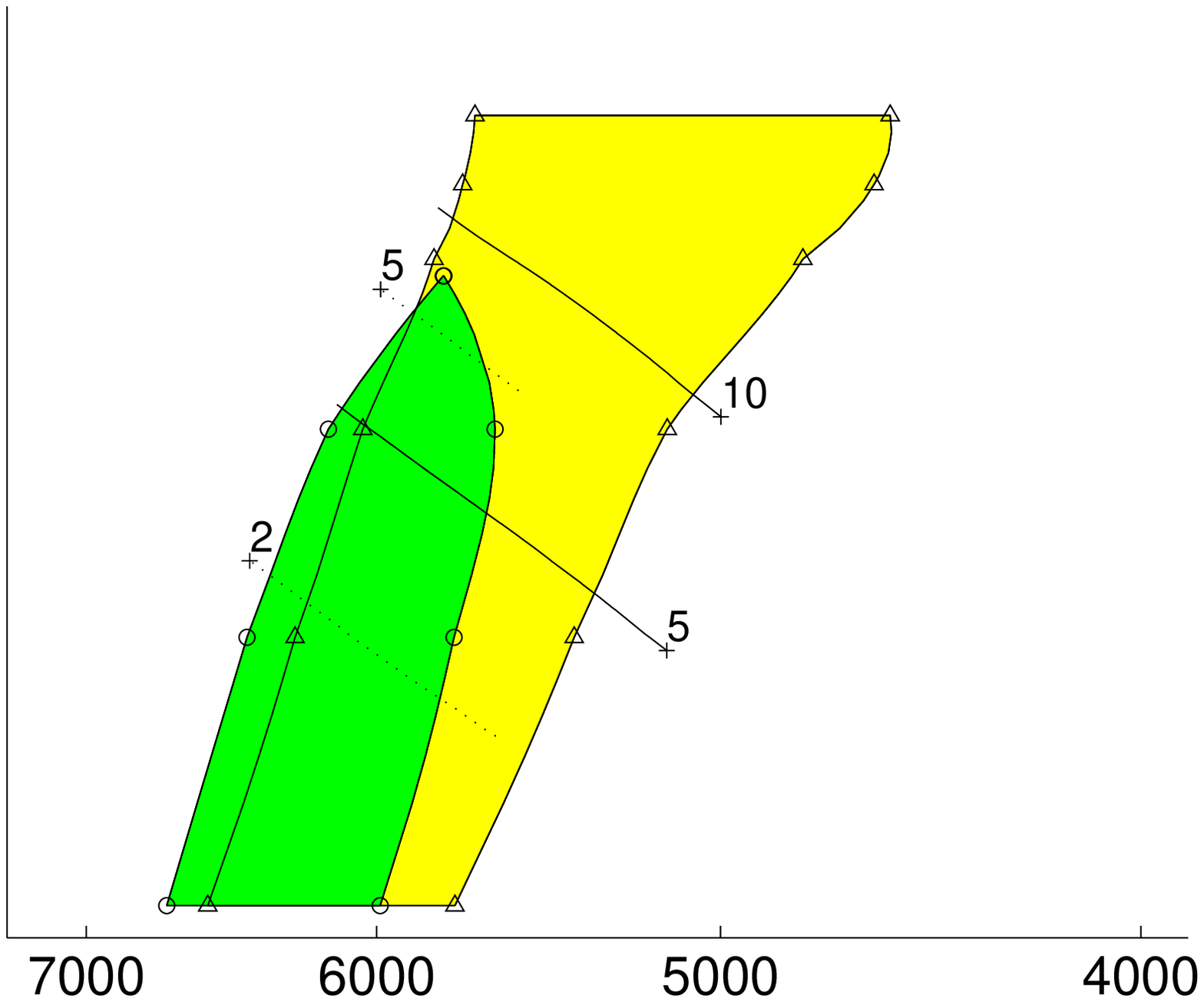}
                       }
 \hfill
 \resizebox{\hsize}{!}{\includegraphics{nullhv.ps}
                       \includegraphics{nullh.ps}
                       \includegraphics{nullh.ps}
                       \includegraphics{nullh.ps}
                       \includegraphics{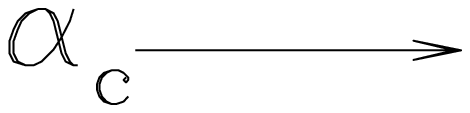}
                       }
 \parbox[b]{\hsize}{
   \caption{
 Linear instability strips as a function of $\alpha_\nu$  and
 $\alpha_c$ with fixed $\alpha_\Lambda$ = 0.375, 
 fundamental IS (light shaded area) and first-overtone IS (dark area).
 {\sl Bottom to top:} $\alpha_\nu$ = 1.33, 1.67, 2.0, 2.33;  
 {\sl left to right:} $\alpha_c$ = 2, 3, 4, 5.
   }
 \label{cis}
 }
 \end{figure*}
 % END FIGURE 14

 Most of the period shift in the TC models as compared to their
radiative equivalents is due to the structural rearrangement that
convection causes in the equilibrium model.  These structurally induced
changes vary from less than 1\% near the blue edge to as much as 15\%
near the red edge, for the very most turbulent models.

The effects of the perturbations of the turbulent energy equation on the
pulsation can also cause a shift in the linear periods (in a way that is
similar to the nonadiabatic temperature perturbations).  These TC shifts
are generally less than one percent when compared to the nonadiabatic
nonturbulent models.

 \section {Instability Strips}

The shape and location of the IS depends on the $\alpha$ parameters that
ultimately we want to determine from the observational constraints.  To
this effect we now calculate Cepheid model sequences with masses in the
range $M$=3 -- 7\Mo.  The composition is taken to be $X=0.70$, $Z=0.02$,
and $L$ is determined from an assumed mass-luminosity relation $\log L =
0.83187 + 3.56 \log M$.  At each $M$ (and $L$) a sequence is calculated
with \Teff\ ranging from $4000$ to $7000$\th K in steps of
$100$--$200$\th K.  At all $M$ and $L(M)$ the \Teff\ dependent
growth-rates functionally resemble those of Fig.~\ref{kmlt}.

 The global picture can be visualized in Fig.~\ref{etasurf}, which shows
a typical surface $\eta_0 = \eta_0(L,$\Teff$)$ (obtained from a cubic
spline interpolation of all the computed data).  The blue and red edges
are shown as contour lines.

We turn now to the morphology of the IS in theoretical HR diagrams as a
function of some of the $\alpha$ parameters.

Fig.~\ref{mltis} shows the results for the mixing length parameter
$\alpha_\Lambda$ (increasing rightward with values 0.288, 0.328, 0.375,
0.390) and $\alpha_\nu$ (increasing vertically with values 1.5, 2.0,
2.5, 3.0), and with $\alpha_c $=3. (the $\alpha_p, \alpha_t, \alpha_s,
\alpha_d$ are kept at their standard values).  The light-shaded regions
form the fundamental IS \th ($\eta_0>0$) and the dark-shaded ones the
first overtone IS \th ($\eta_1>0$).  (The dots represent calculated
models, but some of the sharp features are a result of the interpolation
and of the relatively crude numerical mesh).  One notes that the
sensitivity to $\alpha_\Lambda$ is very large.

Too weak a convective flux parameter (small $\alpha_c$) gives too broad
an IS, both for the fundamental (F) and for the first overtone (O1).
However, a small mixing length has a similar effect.  On the other hand,
too large a convective parameter kills off O1.  The figure shows that
various combinations of parameter values provide approximatively
equivalent IS's, e.g. running approximately from the top middle toward
the bottom right.

Fig.~\ref{cis} shows the morphology of the IS in theoretical HR diagrams
as a function of the convective flux parameter $\alpha_c$ (increasing
from left to right with values 2, 3, 4, 5) and $\alpha_\nu$ (increasing
from the bottom to the top with values 1.0, 1.5, 2.0, 2.5), and with
$\alpha_\Lambda$ = 0.375 (the $\alpha_p, \alpha_s, \alpha_t, \alpha_d$
again are kept at their standard values).  This figure again shows that
various combinations of parameter values provide approximatively
equivalent IS's, e.g. running approximately from the left top toward the
bottom right.

The first overtone IS closes off at large $L$ for some range of $\alpha$
values, which is in agreement with observation.  Thus the overtone
Cepheids with the longest observed periods ($P_1$) give a strong
constraint on the value of the $\alpha$'s (and on the $M$ -- $L$
relation).
 For reference we have drawn iso-period lines $P_0$ = 5 and 10~d (solid
lines) and $P_1$ = 2 and 5\th d (dotted) in Figs.~\ref{mltis} and
\ref{cis}.  Overall, changes in the values of the $\alpha_c$ and
$\alpha_\Lambda$ cause small shifts of the isoperiod lines.  As already
mentioned most of these shifts are due to the structural rearrangement
caused by convection.  That is why $\alpha_\nu$, which does not appear
in the static equilibrium equations, has almost no effect.  The
periods are only slightly affected by convection near the blue edge, but
the pulsation--convection coupling can cause more appreciable changes
near the red edge for the more turbulent models.

We conclude that, just from a visual inspection of these IS diagrams, it
is possible to delineate the acceptable ranges of the parameters
$\alpha_\Lambda$, $\alpha_\nu$ and $\alpha_c$.  Roughly speaking the
parameters with reasonable looking IS's lie on a 2D surface in the
($\alpha_c$, $\alpha_\nu$, $\alpha_\Lambda$) parameter space.  Table~I
quantifies this visual result over a more extensive range of the
$\alpha$ parameters.

\begin{table}

\caption[]{Examples of three parameter combinations which give nearly
equivalent results for the F and first overtone instability strips.
 }
 
 \begin{center}
 \vskip .25cm
 \begin{tabular}{rcrcr}
 \hline\noalign{\smallskip}
 \ \ $\alpha_c$\ & \ \quad &\ \ \ \quad $\alpha_\Lambda$\ \quad &\ \ \
&\ \quad $\alpha_\nu$\ \\
 \noalign{\smallskip}
 \hline
 \noalign{\smallskip} 
 \, 0.5 && 0.375 &&  1.50 \\
 \, 0.5 && 0.500 &&  1.00 \\
 \, 3.0 && 0.150 &&  9.00 \\
 \, 3.0 && 0.328 &&  3.00 \\
 \, 3.0 && 0.420 &&  1.00 \\
 10.0 && 0.100 && 25.00 \\
 10.0 && 0.200 &&  6.00 \\
 \noalign{\smallskip}
 \hline
 \end{tabular}
 \end{center}
 \end{table}

This flexibility in the $\alpha$ values appears fortunate because
ultimately we want not just to fit observed fundamental and first
overtone IS's for the Galaxy, but also for the Magellanic Clouds which
have lower $Z$ values.  In addition, the wealth of observational Cepheid
Fourier decomposition parameters of the (nonlinear) light and radial
velocity curves in these galaxies will need to be fitted.
 
 We have thus narrowed down the parameter space on which more detailed
calibrations will be made that take into account the wealth of
observations.  However, we have not made any IS diagrams showing the
effects of the other four $\alpha$'s because they seem to play a lesser
role, at least at the level of accuracy considered in this parameter
survey.

 \section{Conclusions}

 Purely radiative Cepheid models cannot satisfactorily 
account for all observational data, and this 
has forced us to include turbulence and convection in our Cepheid
models.  We have adopted a simple 1D model diffusion equation for
turbulent convection (TC) that can easily be incorporated into a radial
stellar pulsation code.  Although there is no guarantee that such a
single 1D equation can satisfactorily approximate the effects of
turbulence and convection it is worth exploring to what extent we can
obtain agreement with observation.
 The model TC equations contain several dimensionless, order unity
parameters that are directly related to the physical quantities of the
model that need to be calibrated with the help of observational
astronomical constraints.

 Instead of performing time-consuming fully nonlinear hydrodynamic
computations of the pulsations, we have developed a code to perform an
efficient linear nonadiabatic stability analysis of the models.  This
has allowed us to make an exploration of the sensitivity to the free
parameters of the TC equation, both of the structure of the equilibrium
Cepheid models and of their linear properties.

In agreement with other studies we find that the static Cepheid models
exhibit convection primarily in their H and He partial ionization
regions (PIRs), although in some parameter range a convective zone also
appears in the Fe PIR.  For strongly convective models the separate
convective zones merge into a large one that penetrates up to 320,000\th
K.

The coupling of pulsation and convection also modifies the linear
properties of the pulsational modes, as expected.  It also creates a new
branch of turbulent diffusion modes linear that our stability analysis
establishes to be extremely damped, fortunately for our 1D TC recipe.
These additional modes thus cannot create havoc numerically or
physically. This large damping is consistent with the convective
timescales being very short compared to the periods of the vibrational
of the excited modes which themselves are short compared to the
growth-times.

Our survey shows that the stability of the fundamental and first
overtone modes depends on all the TC parameters, but that it is
dominated by (1) the convective flux, (2) the eddy viscosity, and (3)
the mixing length.  Generally, the effects of eddy viscosity and mixing
length are greater in the higher overtones.

In agreement with other much less detailed TC studies we find that
reasonable $\alpha$ parameter values exist for which IS's of the galactic
Cepheid F and O1 have an acceptable range of temperatures.  In this
paper we have made no effort to narrow down the acceptable range of TC
parameters by using the large amount of observational information.
An application of the TC code to a broader sample of Cepheid
models and its comparison to the large data base of Galactic, and
Magellanic Cloud Cepheids is in progress, as is an extension to RR~Lyrae
stars.

 \section{Acknowledgements}
 This work has been supported by NSF (grants AST95--28338 and INT94--15868).
 P.Y. gratefully acknowledges helpful conversations with Neil Balmforth and
 Ed Spiegel.

\end{document}